\documentclass[reprint,pra]{revtex4-1}
\usepackage{amsmath}
\usepackage{amsfonts}
\usepackage{color}
\usepackage[utf8]{inputenc}
\usepackage[dvips]{graphicx}
\usepackage{indentfirst}
\usepackage{fancyhdr}
\usepackage{float}
\usepackage{subfigure}
\begin{document}
\title{\textbf{Trap effects and continuum limit of the Hubbard model in the presence of a harmonic potential}}
\author{Davide Nigro}
\affiliation{Dip. di Fisica dell’Università di Pisa, Largo Pontecorvo 3, I-56127 Pisa, Italy}
\begin{abstract}
We give a prescription to perform the continuum limit of the $d$-dimensional Hubbard model in the presence of a harmonic trap at zero temperature. We perform the continuum limit at fixed number of particles. In $d\geq3$ the lattice system of spin-1/2 particles is mapped into a non-interacting two-component Fermi gas in a harmonic trap. In $d=1$ and $d=2$ the particles with opposite spin interact via a Dirac delta interaction. We show that the properties of this continuum limit can be put in correspondence with those derived applying the Trap-Size scaling (TSS) formalism to the confined Hubbard model in the so called Dilute Regime (fixed number of particles and weak confinement). The correspondence in $d=1$ and $d=2$ has been tested comparing the numerical results obtained for lattice system with those of the continuum limit in the case of two-particle and in absence of spin-polarization ($N=2$,$N_{\uparrow}=N_{\downarrow}=1$).
\end{abstract}
\maketitle
\section{Introduction}\label{sec:intro}
During the last few decades cold atoms systems \cite{Intro1,Intro2}, have been extensively used in many fields of research and have played a key role in particular in the understanding of quantum many-body systems. All these results are related to the impressive progress that has been made in manipulating these many-body systems. 
Nowadays using Lasers it is possible to cool atoms down at very low temperature and to achieve regimes at which only fluctuations due to principles of quantum mechanics are relevant to study the many-body system behaviour. In addition, using Lasers it is possible to tune two-body interactions \cite{Intro4} and moreover it is possible to realize spatially periodic potentials in which cooled neutral atoms congregate in structures resembling usual crystal lattices. These structures are called \emph{optical lattices}. The main difference between optical and ordinary lattices is that all the lattice features are not fixed , but only depend on the experimental setup: tuning Laser wavelength and intensity and modifying the number of counter-propagating beams respectively, one can  change the interatomic distance (\emph{lattice spacing}), the interaction strength (\emph{lattice depth}) and also the \emph{lattice geometry} \cite{Intro3}.\newline
An important feature of all the experiments with cold atoms on optical lattices is the presence of an external space-dependent confining potential. Such a kind of potential is usually harmonically-shaped \cite{Intro5,Intro6} and it directly couples with the particle density of the lattice system, leading to inhomogeneities which give rise to a wide variety interesting physical phenomena (for example the coexistence of many quantum-phases \cite{Exot1}, the presence of an inhomogeneous \emph{Crossover} \cite{Exot2}, etc).\newline 
In a recent paper several of the most interesting features developed by lattice systems in the presence harmonic traps have been considered in details  \cite{TSS4}. There the authors use the Trap-Size Scaling (TSS) formalism and numerical techniques to investigate the properties developed in $d$-dimensional lattice system of spin-$1/2$ particles described by the Hubbard model in the presence of a harmonic trap (see next Section for details). 
While considering the one dimensional problem at very low temperature, fixed number of particles and weak confining potential, they argue about the existence of a one-to-one correspondence between the ground-state properties of the lattice problem and those of the Gaudin-Yang model \cite{Gaudin,Yang} in the presence of a harmonic trap.\\

In this paper we show that this correspondence can be analytically derived performing the proper continuum limit of the lattice problem. We do this for an arbitrary number of particles $N$ and in arbitrary dimension $d$. It turns out that this continuum limit strongly depends on the dimension $d$ and gives results coherent with those obtained by using the Renormalization-Group techniques in \cite{TSS4}. The validity of this correspondence has been tested explicitly by considering what happens to the system in the simplest possible configuration, that is the unpolarized two-body problem ($N=2$ and $N_{\uparrow}=N_{\downarrow}=1$) where both analytical and numerical calculations can be carried out easily.\\ 

The remaining of the paper is organized as follows. In Sec. \ref{sec:hamiltonian model} we introduce the hamiltonian model and we define also the regime of interest for the present work, that is the Dilute Regime. In Sec. \ref{sec:DRandRG} we report the ideas discussed in Ref. \cite{TSS4}. Here we show how it is possible to keep into account the effects introduced by an external confining potential within the TSS formalism.
In Sec. \ref{sec:regularization} we deduce the continuum limit of the Hubbard model in presence of a harmonic coupling and we discuss how this continuum limit can be put in correspondence with the TSS formalism. We refer to this correspondence  by saying ``the \emph{Correspondence Hypothesis}".\\
In Sec. \ref{sec:twobodyproblem} we test the correspondence hypothesis by considering the simplest non-trivial configuration for the system in analysis, that is the two-body unpolarized problem in $d=1$ and $d=2$. For $d\geq 3$ both the continuum limit and the trap-Size Scaling formalism prescribe a trivial behaviour of the lattice system, that is the behaviour of a two-component non-interacting Fermi gas in a harmonic trap.\\
Finally, in Sec. \ref{sec:conclusions} we summarize our main results and draw our conclusions.\\



\section{The Hamiltonian model $\&$ the Dilute Regime}\label{sec:hamiltonian model}
A system of spin-$\frac{1}{2}$ interacting fermions on a $d$-dimensional lattice can be described using the well-known Hubbard Hamiltonian
\begin{equation}\label{eq:HubbardModel}
\mathcal{H}_0=-t\sum_{\sigma}\sum_{\left\langle \textbf{x},\,\textbf{y}\right\rangle}\left(c^{\dagger}_{\textbf{x},\,\sigma}c_{\textbf{y},\,\sigma}+h.c.\right)+U\sum_{\textbf{x}}n_{\textbf{x},\,\uparrow}n_{\textbf{x},\,\downarrow}
\end{equation}
where $\textbf{x}=\left(x_{1},x_{2},\cdots,x_{d}\right)$ denotes a site on the $d$-dimensional lattice, $t$ and $U$ are respectively the hopping and the on-site coupling constant, $\sigma=\uparrow,\,\downarrow$ is the spin label, $\left\langle \cdot\,,\,\cdot\right\rangle$ is the summation over first neighbour sites, $c^{(\dagger)}_{\textbf{x},\,\sigma}$ is the annihilation (creation) operator for a particle with spin $\sigma$ on the site $\textbf{x}$ and $n_{\textbf{x},\,\sigma}=c^{\dagger}_{\textbf{x},\,\sigma}c_{\textbf{x},\,\sigma}$ is the number operator.\\
The presence of an isotropic harmonic potential coupled to the particle density can be taken into account by adding the following Hamiltonian term to the Hubbard model in Eq.(\ref{eq:HubbardModel})
\begin{equation}\label{eq:TrapTerm}
\mathcal{H}_{t}=\sum_{\sigma}\sum_{\textbf{x}}\frac{1}{2}v^{2}\vert\vert \textbf{x}\vert\vert^{2}n_{\textbf{x},\,\sigma}
\end{equation}
where $v$ is the trap-intensity and $\vert\vert \textbf{x}\vert\vert^{2}=\sum_{\alpha=1}^{d}x_{\alpha}^{2}$ quantifies the distance between the site $\textbf{x}$ and the centre of the lattice structure.\\
The total Hamiltonian now reads
\begin{equation}\label{eq:total hamiltonian}
\mathcal{H}=\mathcal{H}_{0}+\mathcal{H}_{t}
\end{equation}
In the following sections we refer to the Hamiltonian in Eq.(\ref{eq:total hamiltonian}) by saying \emph{confined Hubbard model}.
The confining potential introduces inhomogeneities that can be characterized in terms a new characteristic length scale, $l$, called \emph{trap-size} which is defined as follows
\begin{equation}\label{eq:trapsizedef}
l=\frac{\sqrt{2t}}{v}
\end{equation}
The definition in Eq.(\ref{eq:trapsizedef}) naturally arises when trying to define the analogue of the thermodynamic limit in the presence of a harmonic confinement as discussed in Ref.\cite{sachdevttl} for a system of confined bosons.\\
In terms of the trap-size $l$ and the total mean-number of particles
\begin{equation}\label{eq:number of particles}
N=\left\langle\sum_{\textbf{x},\,\sigma}n_{\textbf{x},\,\sigma}\right\rangle
\end{equation}
it is possible to define two different regimes related to the ratio
\begin{equation}\label{eq:ratiodef}
\rho=\frac{N}{l^{d}}
\end{equation}
which quantifies the mean-number of particles confined inside the $d$-dimensional ``trap-volume" $l^{d}$.
The main difference between these two regimes is the way in which the number of particles $N$ changes while increasing the trap-size $l$.\\
The first regime is called \emph{Trap} \emph{Thermodynamic} \emph{Limit} (TTL). The TTL is the asymptotic regime obtained performing both the large $N$ and the large $l$ limits while keeping $\rho$ constant
\begin{equation}\label{eq:TTLdef}
\textbf{TTL :}\quad N\to +\infty,\,l\to +\infty,\,\rho=\mbox{constant}
\end{equation}
The other one is called \emph{Dilute} \emph{Regime} (DR). The DR, which is the regime of interest in the present work, is the asymptotic condition obtained performing the large $l$ limit at fixed number of particles $N$
\begin{equation}\label{eq:DRdef}
\textbf{DR :}\quad N=\mbox{constant},\,l\to +\infty,\,\rho\to 0
\end{equation}
As discussed in details for the Bose-Hubbard model in \cite{tssCanonicalGCanonical} and for the fermionic case in \cite{TSS4}, the properties shown by the lattice system in the two regimes introduced above can be studied by considering the large-$l$ behaviour of the \emph{Grand Canonical} (GC) counterpart of the Hamiltonian in Eq.(\ref{eq:total hamiltonian}) within the  Trap-Size Scaling (TSS) formalism (\emph{i.e.} Renormalization Group formalism in the presence of spatial inhomogeneities). In the GC formalism one modifies Eq.(\ref{eq:total hamiltonian}) by adding a term proportional to the chemical potential $\mu$. The particular value of the parameter $\mu$ specifies completely the large-$l$ regime under study. In other words the choice of $\mu$ specifies the way in which the number of particles $N$ behaves as the trap-size $l$ increases.\\
In the following section we review how this correspondence can be used within the TSS formalism to characterize the DR properties of the hamiltonian model in Eq.(\ref{eq:total hamiltonian}).\\ 
\section{The trap-size scaling formalism in the Dilute Regime}\label{sec:DRandRG}
The GC counterpart of the hamiltonian model in Eq.(\ref{eq:total hamiltonian}) reads
\begin{equation}\label{eq:grandcanonical hubbard}
\mathcal{H}_{GCC}=\mathcal{H}_{0}+\mathcal{H}_{t}+\mathcal{H}_{\mu},\\
\end{equation}
where
\begin{equation}\label{eq:chemicalpotential term}
\mathcal{H}_{\mu}=-\mu\sum_{\textbf{x},\,\sigma}n_{\textbf{x},\,\sigma}.
\end{equation}
As discussed in \cite{TSS4}, in the DR the hamiltonian term $\mathcal{H}_t$ can be treated as a weak perturbation to $\mathcal{H}_{GC}=\mathcal{H}_0 + \mathcal{H}_{\mu}$. This means that to characterize all the interesting features developed by the system due to the external confinement, that is the scaling properties of the correlation functions, one characterizes first the critical features of the hamiltonian without confinement and then consider the effects introduced by the presence of the external potential, that is the existence of a finite correlation length related to the trap-size $l$. The method described above is the basis of the TSS formalism, which has been extensively used during the last few decades to analyze the properties of confined lattice systems.\\
The critical properties of $\mathcal{H}_{GC}$ can be deduced by the studying the following Quantum Field Theory \cite{Sachdev,path1} (we set $\hbar=1$,$k_B=1$)
\begin{equation}\label{eq:quantumtheory}
\begin{split}
&\mathcal{Z}=\int \mathcal{D}\psi_{\sigma}^{*} \mathcal{D}\psi_{\sigma}\mbox{exp}\left(-\int_{0}^{1/T}\mbox{d}\tau\mbox{d}^{d}x\,\mathcal{L} \right),\\
&\mathcal{L}=\sum_{\sigma}\left[\psi^{*}_{\sigma}\frac{\partial}{\partial_{\tau}}\psi_{\sigma}+\frac{1}{2M}\vert\nabla\psi_{\sigma}\vert^{2}-\bar{\mu}\vert\psi_{\sigma}\vert^{2}\right]+\\
&+u\psi_{\uparrow}^{*}\psi_{\downarrow}^{*}\psi_{\downarrow}\psi_{\uparrow}\\
\end{split}
\end{equation}
where $T$ is the temperature of the system and $\psi_{\sigma}=\psi_{\sigma}(\textbf{x},\,\tau)$ and $\psi^{*}_{\sigma}=\psi_{\sigma}(\textbf{x},\,\tau)$ are the Grassmann fields associated to the creation and annihilation operators in Eq.(\ref{eq:grandcanonical hubbard}), $M$ denotes the mass of the fields, $u$ is the many-body interaction strength and $\bar{\mu}$ is the chemical potential.\\
The quantum theory in Eq.(\ref{eq:quantumtheory}) has a Gaussian fixed-point in correspondence of
\begin{equation}\label{eq:dilutefixedpointquantumtheory}
\bar{\mu}=0,\quad T=0,\quad u=0
\end{equation}
which corresponds to the following point in parameter space for the hamiltonian $\mathcal{H}_{GC}$
\begin{equation}\label{eq:dilutefixedpointhubbard}
\mu=-2t,\quad T=0,\quad U=0.
\end{equation}
This fixed-point prescribes the scaling behaviour for the theory in Eq.(\ref{eq:quantumtheory}) in correspondence of a \emph{Metal-Vacuum} \emph{Quantum Phase transition} (QPT). It is interesting to observe that this QPT is characterized by a low mean-occupation, which is physically consistent with the low mean-density condition which characterises the DR.\\
Since the fixed-point in Eq.(\ref{eq:dilutefixedpointquantumtheory}) is Gaussian, the scaling behaviour of all the variables entering in Eq.(\ref{eq:quantumtheory}) 
can be determined easily. Under a scaling transformation of the form
\begin{equation}\label{eq:scaling transformation}
\textbf{x}\to\textbf{x}'=\frac{\textbf{x}}{b}
\end{equation} 
we have that the variables entering in the lagrangian $\mathcal{L}$ transform this way
\begin{equation}
\psi_{\sigma}\to\psi'_{\sigma}=b^{y_{c}}\psi_{\sigma},\,\, u\to u'=b^{y_{u}}u,\,\,\bar{\mu}\to\bar{\mu}'=b^{y_{\bar{\mu}}}\bar{\mu},
\end{equation}
where $y_{c}$,  $y_{\bar{\mu}}$ and $y_{u}$ are the Renormalization Group (RG) exponents associated respectively  to the \emph{field} $\psi_{\sigma}$, to the \emph{chemical potential} $\bar{\mu}$ and to the \emph{many-body interaction} $u$.
By performing a simple dimensional analysis one finds the following values for the RG-exponents and for the \emph{dynamic critical exponent} z \cite{Sachdev,CardyRG}
\begin{equation}\label{eq:rg dimensions}
y_{c}=\frac{d}{2},\quad y_{\bar{\mu}}=2,\quad y_{u}=2-d
\end{equation}
\begin{equation}\label{eq:dynamic exponent}
z=2
\end{equation}
Let us now consider the effects of the confinement.
In this framework the confining term can be introduced using the following contribution
\begin{equation}\label{eq:traptermfieldtheory}
\mathcal{L}_{v}=\frac{1}{2}v^{2}\textbf{x}^{2}\sum_{\sigma}\vert\psi_{\sigma}\vert^{2}
\end{equation}
From Eq.(\ref{eq:rg dimensions}) and Eq.(\ref{eq:dynamic exponent}), proceeding in the same way of Refs. \cite{tssCanonicalGCanonical,TSS4,TSS3} the RG-exponent associated with the trap-intensity $v$ is\\
\begin{equation}\label{eq:trap exponent}
y_{v}=2.
\end{equation}
The confining potential introduces a new \emph{characteristic length-scale} $\xi$ that is related to the trap-size by the following power-law dependence
\begin{equation}\label{eq:characteristic length}
\xi \sim l^{\theta},
\end{equation}
being
\begin{equation}\label{eq:theta parameter}
\theta=1/y_{v}=1/2
\end{equation}
The presence of this characteristic length-scale modifies the correlation between particles, leading to a non-trivial scaling behaviour of all the expectation values called \emph{Trap-Size Scaling} (TSS). 
Let us consider a generic operator $\mathcal{O}\left(\textbf{x};v,U\right)$ with RG-exponent $y_{o}$ and let us consider also its n-points correlation function
\begin{equation}\label{eq:n point correlatione function}
\mathcal{W}\left(\textbf{x}_1,\cdots,\textbf{x}_n;v,U \right)\equiv\langle \mathcal{O}\left(\textbf{x}_1; v,U\right),\cdots, \mathcal{O}\left(\textbf{x}_n; v,U\right)\rangle
\end{equation}
where $\langle \cdot \rangle$ denotes the expectation value on the ground-state configuration. We obtain the TSS behaviour of the correlation function in Eq.(\ref{eq:n point correlatione function}) when considering a scaling transformation with parameter $b=l^{\theta}$.
In this case a scaling transformation of the lattice structure of the form $\textbf{x}\to\textbf{x}'=\textbf{x}l^{-\theta}$ induces the following scaling transformation of $\mathcal{W}\left(\textbf{x}_1,\cdots,\textbf{x}_n;v,U \right)$
\begin{equation}\label{eq:TSS npoint correlation function}
\begin{split}
\mathcal{W}\left(\textbf{x}_1,\cdots,\textbf{x}_n;\right. & v, \left. U \right)=\mathcal{W}\left(\textbf{x}'_1,\cdots,\textbf{x}'_n; v', U' \right)=\quad\\
&=l^{-\theta y_{\mathcal{W}}}\mathcal{W}\left(\frac{\textbf{x}_1}{l^{\theta}},\cdots,\frac{\textbf{x}_n}{l^{\theta}};1, Ul^{\theta y_{u}}\right)\equiv\\
&\equiv l^{-\theta y_{\mathcal{W}}} \mathcal{F}\left(\textbf{X}_1',\cdots,\textbf{X}_n';U_r\right),
\end{split}
\end{equation}
where $y_{\mathcal{W}}=n y_{O}$ denotes the RG-exponent of the n-point correlation function in Eq.(\ref{eq:n point correlatione function}), $\mathcal{F}\left(\textbf{X}_1,\cdots,\textbf{X}_n;U_r\right)$ is the \emph{scaling function} of Eq.(\ref{eq:n point correlatione function}) and $U_r\equiv Ul^{\theta y_{u}}$. We have to stress that the equivalence in Eq.(\ref{eq:TSS npoint correlation function}) is true only in the large-$l$ limit and in general for finite-$l$ there are corrections which slightly modifies the scaling behaviour of the correlation functions.\\
For our purposes it is interesting to observe that the on-site coupling constant $U$ shows a RG-behaviour which depends on the dimension $d$. In $d=1$, where the on-site coupling $U$ is \textbf{relevant}, the scaling functions are expected to depend strongly on this parameter. In $d=2$, where $U$ is \textbf{marginal}, one expects to observe only a residual dependence of the scaling functions on the on-site interaction. For $d \geq 3$ instead, where this parameter becomes \textbf{irrelevant}, the TSS behaviour is expected to match the behaviour of a two-component Fermi gas in the presence of a harmonic trap.\\
The TSS Ansatz in Eq.(\ref{eq:TSS npoint correlation function}) has been extensively tested and discussed in \cite{TSS4} for the one-dimensional case. Numerical results reported therein show that while increasing the value of the trap-size $l$ at fixed $U_r=Ul^{\theta}$ data seem to converge to non-trivial curves which depends on the particular value of the parameter $U_r$. In particular results obtained for the mean-occupation $\langle n_{\textbf{x}}\rangle =\sum_{\sigma}\langle n_{\textbf{x},\,\sigma}\rangle $ suggest the existence of a correspondence between the asymptotic scaling properties of the one-dimensional confined Hubbard model (at fixed $N$ and $T=0$) and the Gaudin-Yang model \cite{Gaudin,Yang} in presence of an external harmonic potential.\\
In the next section we show that there is not only a mapping between the asymptotic properties of the lattice system in the DR and the Gaudin-Yang model with an external confining potential in $d=1$, but also that is possible to map the lattice theory into a continuous one in any $d$.\\

\section{Continuum limit of the confined Hubbard model $\&$ TSS properties}\label{sec:regularization}
In this section we show how to introduce a continuous theory equivalent to the confined Hubbard model in the large-$l$ regimes discussed in the previous sections and we explicitly show how this limit is related to the results obtained within the TSS formalism.\\
\subsection{The continuum limit in the presence of a harmonic trap}\label{subsec:continuumlimit}
The idea at the basis of the method discussed here is quite simple and it is the following. As said before the presence of the confining potential introduces a new length scale $l$. To understand how this new length scale affects the system, we have to compare $l$ to the lattice spacing $a$, which is the intrinsic length scale of the lattice theory. In fact, the system properties do not depend on these two lengths separately, but they are determined by the ratio $a/l$. This consideration tells us that the case in which we send $l$ to infinity keeping fixed $a$ (which at fixed number of particles defines the DR) and the case in which we send $a$ to zero keeping constant the trap-size $l$ must show the same physical properties. Therefore, by performing a small-$a$ analysis of the trapped Hubbard model at fixed number of particles and fixed $l$, it is possible to introduce a continuous theory (Continuum limit) by means of which we can describe the properties developed by the lattice problem in the DR.\\
The Continuum limit of the lattice theory in the DR can be introduced by first perfoming a small-$a$ expansion of the hamiltonian model in Eq.(\ref{eq:grandcanonical hubbard}) in correspondence of $\mu=-2t$ which is the value prescribed by the RG analysis (see Eq.(\ref{eq:dilutefixedpointhubbard})) and then by taking the following limit 
\begin{equation}\label{eq:regularization}
\mathcal{H}_{c}^{(d)}=\lim_{a\to 0}\frac{1}{a^2}\left[\mathcal{H}_{GCC}\vert_{\mu=-2t}+t(d-1)N\right]
\end{equation}
where $N=\sum_{\textbf{x},\,\sigma}n_{\textbf{x},\,\sigma}$. Here we consider the DR, but we expect this procedure to be quite general: if one is interested in the system properties in correspondence of an another large-$l$ regime, one has simply to choose properly the value of the chemical potential $\mu$ and to modify multiplicative constant in front of $N$ to cancel all the contributions proportional to the number of particles.\\
If one defines the following field operators at fixed $\textbf{x}=\textbf{j}\,a$ \cite{ESSLER}
\begin{equation}\label{eq:pseudo field operators}
\Psi^{\dagger}_{\sigma}\left(\textbf{x}\right)=\left.\lim_{a\to 0}\frac{C_{\textbf{j},\,\sigma}^{\dagger}}{a^{d/2}}\right|_{\textbf{x}=\textbf{j} a},\quad
\Psi_{\sigma}\left(\textbf{x}\right)=\left.\lim_{a\to 0}\frac{C_{\textbf{j},\,\sigma}}{a^{d/2}}\right|_{\textbf{x}=\textbf{j} a}
\end{equation}
the result of the limit in Eq.(\ref{eq:regularization}) is the following
\begin{equation}\label{eq: regularaized model}
\begin{split}
&\quad\quad\quad\quad\mathcal{H}_{c}^{(d)}=-t\sum_{\sigma}\int d^dx\,\Psi_{\sigma}^{\dagger}(\textbf{x})\nabla^2_d\Psi_{\sigma}(\textbf{x})+\\
&+g_d \int d^dx\int d^dy \Psi^{\dagger}_{\uparrow}(\textbf{x})\Psi^{\dagger}_{\downarrow}(\textbf{y})\delta(\textbf{x}-\textbf{y})\Psi_{\downarrow}(\textbf{y})\Psi_{\uparrow}(\textbf{x})+\\
&\quad\quad\quad\quad+\frac{v^{2}_{c}}{2}\sum_{\sigma}\int d^{d}x\vert\vert\textbf{x}\vert\vert^{2}\Psi^{\dagger}_{\sigma}(\textbf{x})\Psi_{\sigma}(\textbf{x}),\\
\end{split}
\end{equation}
where $\nabla^2_d$ is the \emph{Laplace operator} in dimension $d$ and 
\begin{equation}\label{eq:couplingcontinuumlimit}
v_{c}=\frac{v}{a^{2}}\,\quad g_d=Ua^{d-2}.
\end{equation}
If we now use the well-known relations between first and second quantization formalism \cite{Landau}, by setting 
\begin{equation}\label{eq:sustitutions}
t=\frac{\hbar^2}{2m},\quad v_{c}^{2}=m\omega^2,
\end{equation}
the model in Eq.(\ref{eq: regularaized model}) is equivalent to following $d$-dimensional continuous Hamiltonian model
\begin{equation}\label{eq:continuum model 2}
\mathcal{H}_{c}^{(d)}=\sum_{i=1}^{N}\left(\frac{\textbf{p}_{i}^2}{2m}+\frac{1}{2}m\omega^2\textbf{x}_{i}^{2}\right) +g_d\sum_{i=1}^{N_{\uparrow}}\sum_{j=1}^{N_{\downarrow}}\delta(\textbf{x}_i-\textbf{x}_j).
\end{equation}
where $N_{\sigma}$ is the number of particles with spin polarization $\sigma$ ($\sigma=\uparrow,\,\downarrow$).\\

The hamiltonian in Eq.(\ref{eq:continuum model 2}) describes the dynamics of a system of $N$ spin-$\frac{1}{2}$ fermions into an external harmonic potential. The presence of the local delta interaction depends on the dimension $d$. This fact is coherent with the results obtained by using the RG formalism. This becomes clearer if we explicit consider $g_d$ in $d=1$, $d=2$ and $d\geq 3$.\\
According to Eq.(\ref{eq:couplingcontinuumlimit}) we have that
\begin{equation}
\left\lbrace
\begin{split}
& g_1=Ua^{-1},\,\mbox{in}\,d=1\\
& g_2=U,\,\,\,\mbox{in}\,d=2\\
& g_d=Ua^{d-2},\,\,\,\mbox{in}\,d\geq 3.\\
\end{split}
\right.
\end{equation}
In $d=1$, where the the on-site parameter is a \emph{relevant} variable, we have that the continuum limit of the lattice model is a strong interacting many-body theory: for an arbitrary $\vert U\vert\neq 0$, it does not matter how small it is, the lattice theory is mapped into an interacting model with $\vert g_1\vert\neq 0$. This means that in $d=1$ the properties of the continuum limit are strongly affected by the presence of an arbitrary small interparticle interaction in the lattice model. In $d=2$, where the RG behaviour of the on-site parameter $U$ is \emph{marginal}, we have that $g_2=U$. This means that the continuum limit preserves the nature and the role played by the on-site coupling in the original theory. In $d\geq 3$ everything changes: since $d-2>0$, the continuum limit in $d\geq 3$ is always a non-interacting theory, that is $g_d$ is always zero. This is coherent with the \emph{irrelevant} RG behaviour found for the $U$ in dimensions higher than $d=2$.\\
If we now use the definition of the trap-size $l$ in Eq.(\ref{eq:trapsizedef}) and Eqs.(\ref{eq:couplingcontinuumlimit}) and (\ref{eq:sustitutions}), we can ``eliminate" the lattice spacing $a$ and we can rephrase the correspondence between $g_d$ and $U$ in terms of the trap-size $l$ and the others parameters entering in the continuum limit model. The relation is the following:
\begin{equation}\label{eq: g-U}
g_{d}=U\left(\frac{\hbar}{m\omega}\right)^{\frac{d-2}{2}}l^{\frac{2-d}{2}}.
\end{equation}
If we neglect the $d\geq 3$ case that is trivial we have that
\begin{equation}\label{eq: g-U dim1}
g_{1}=U_r\left(\frac{m\omega}{\hbar}\right)^{1/2},
\end{equation}
and we have 
\begin{equation}\label{eq: g-U dim2}
g_{2}=U,
\end{equation}
where we used the definition of the rescaled coupling $U_r=Ul^{1/2}$.\\
The relation in Eq.(\ref{eq: g-U}) is the main result of this work. In the following we will essentially discuss the correspondence prescribed by the Eq.(\ref{eq: g-U}) by considering the scaling properties developed by the mean-occupation of the lattice problem.\\ 
\subsection{The correspondence hypothesis: continuum limit $\&$ TSS formalism}\label{sec:correspondencehypothesis}
To test the validity of the correspondence discussed above, the idea is to compare the TSS behaviour of a given observable $\mathcal{O}(\textbf{x}_1,\,\cdots,\,\textbf{x}_n)$ obtained for the lattice problem with its analogue in first-quantization formalism, that is the mean-value of the corresponding operator on the state representing the ground-state configuration $\vert \Psi_{GS};\,g_d \rangle$.\\
In this work we consider several different one-point and two-point observables. In particular in $d=1$ we consider the mean-occupation
\begin{equation}\label{eq:meanoccupationobs}
\rho(\textbf{x};\,v,\,U)=\langle n_{\textbf{x}}\rangle=\sum_{\sigma=\uparrow,\,\downarrow}\langle n_{\sigma,\,\textbf{x}}\rangle,
\end{equation}
the double-occupancy
\begin{equation}\label{eq:doubleoccupancy}
d_0(\textbf{x};\,v,\,U)=\langle n_{\uparrow,\,\textbf{x}} n_{\downarrow,\,\textbf{x}}\rangle,
\end{equation}
the pair-correlation
\begin{equation}\label{eq:paircorrelation}
P(\textbf{x},\textbf{y};\,v,\,U)=\langle C_{\uparrow,\,\textbf{x}}^{\dagger}C_{\downarrow,\,\textbf{y}}^{\dagger}C_{\downarrow,\,\textbf{y}}C_{\uparrow,\,\textbf{y}} + h.c. \rangle,
\end{equation}
the one-particle correlation
\begin{equation}\label{eq:oneparticlecorrelation}
C(\textbf{x},\textbf{y};\,v,\,U)=\sum_{\sigma=\uparrow,\,\downarrow}\langle C_{\sigma,\,\textbf{x}}^{\dagger}C_{\sigma,\,\textbf{y}} + h.c.\rangle,
\end{equation}
and the two connected density-density correlations
\begin{center}
\begin{equation}\label{eq:Gdd}
G(\textbf{x},\textbf{y};\,v,\,U)=\langle n_{\textbf{x}} n_{\textbf{x}}\rangle-\langle n_{\textbf{x}}\rangle\langle n_{\textbf{y}}\rangle,
\end{equation}
\end{center}
\begin{equation}\label{eq:magnetizaion}
M(\textbf{x},\textbf{y};\,v,\,U)=\langle n_{\uparrow,\,\textbf{x}} n_{\downarrow,\,\textbf{x}}\rangle\langle n_{\uparrow,\,\textbf{x}}\rangle\langle n_{\downarrow,\,\textbf{y}}\rangle.
\end{equation}
\newline
In $d=2$ instead we only consider the scaling behavior of the mean-density.\\
According to the analysis performed in Ref.\cite{TSS4} the scaling behaviour of the oservables introduced above under a scaling transformation of the lattice structure of the form $\textbf{x}\to \textbf{X}=\textbf{x}/l^{\theta}$ can be cast in the following form
\begin{equation}\label{eq:scalingrho}
\rho\left(\textbf{x};\,v,\,U\right)\approx l^{-d\theta} \mathcal{R}\left(\textbf{X};\,U_{r}\right),
\end{equation}
\begin{equation}\label{eq:scalingdouble}
d_0\left(\textbf{x};\,v,\,U\right)\approx l^{-2d\theta} \mathcal{D}\left(\textbf{X};\,U_{r}\right),
\end{equation}
\begin{equation}\label{eq:scalingpair}
P\left(\textbf{x},\,\textbf{y};\,v,\,U\right)\approx l^{-2d\theta} \mathcal{P}\left(\textbf{X},\,\textbf{Y};\,U_{r}\right),
\end{equation}
\begin{equation}\label{eq:scalingoneparticle}
C\left(\textbf{x},\,\textbf{y};\,v,\,U\right)\approx l^{-d\theta} \mathcal{C}\left(\textbf{X},\,\textbf{Y};\,U_{r}\right),
\end{equation}
\begin{equation}\label{eq:scalingdensitydensity}
G\left(\textbf{x},\,\textbf{y};\,v,\,U\right)\approx l^{-2d\theta} \mathcal{G}\left(\textbf{X},\,\textbf{Y};\,U_{r}\right),
\end{equation}
\begin{equation}\label{eq:scalingmagnetization}
M\left(\textbf{x},\,\textbf{y};\,v,\,U\right)\approx l^{-2d\theta} \mathcal{M}\left(\textbf{X},\,\textbf{Y};\,U_{r}\right).
\end{equation}

In the DR and for a generic number of particles $N=N_{\uparrow}+N_{\downarrow}$ we expect to have the following expressions for the scaling functions:

\begin{equation}\label{eq:funzionediscalarho}
\mathcal{R}(\textbf{X},\,U_r)= \rho_{GS}\left(\textbf{X};\,g_d\right),
\end{equation}
\begin{widetext}
\begin{equation}\label{eq:funzionediscaladouble}
\mathcal{D}\left(\textbf{X};\,U_{r}\right)= N_{\uparrow}N_{\downarrow}\int\prod_{i=2}^{N_{\uparrow}}\prod_{j=2}^{N_{\downarrow}}\mbox{d}\textbf{x}_i \mbox{d}\textbf{y}_j\left\vert\Psi_{GS}\left(\textbf{X},\textbf{x}_2,\cdots,\textbf{x}_{N_{\uparrow}},\textbf{X},\textbf{y}_2,\cdots,\textbf{y}_{N_{\downarrow}};\,g_d\right)\right\vert^2,
\end{equation} 
\begin{equation}\label{eq:funzionediscalaoneparticle}
\begin{split}
&\mathcal{C}\left(\textbf{X},\,\textbf{Y};\,U_{r}\right)=N\int\prod_{i=2}^{N_{\uparrow}}\prod_{j=1}^{N_{\downarrow}}\mbox{d}\textbf{x}_i \mbox{d}\textbf{y}_j\left[\Psi^{*}_{GS}\left(\textbf{X},\textbf{x}_2,\cdots,\textbf{x}_{N_{\uparrow}},\textbf{y}_1,\cdots;\,g_d\right)\Psi_{GS}\left(\textbf{Y},\textbf{x}_2,\cdots,\textbf{x}_{N_{\uparrow}},\textbf{y}_1,\cdots;\,g_d\right)+c.c.\right],
\end{split}
\end{equation}
\begin{equation}\label{eq:funzionediscalapair}
\begin{split}
&\mathcal{P}(\textbf{x},\textbf{y};U_{r})=N_{\uparrow}N_{\downarrow}\int\prod_{i=2}^{N_{\uparrow}}\prod_{j=2}^{N_{\downarrow}}\mbox{d}\textbf{x}_i \mbox{d}\textbf{y}_j
\left[\Psi^{*}_{GS}\left(\textbf{X},\textbf{x}_2,\cdots,\textbf{X},\textbf{y}_2,\cdots;\,g_d\right)\Psi_{GS}\left(\textbf{Y},\textbf{x}_2,\cdots,\textbf{Y},\textbf{y}_2,\cdots;\,g_d\right)+ c.c. \right],
\end{split}
\end{equation}
\begin{equation}\label{eq:funzionediscaladensity}
\begin{split}
\mathcal{G}(\textbf{x},\textbf{y};\,U_r)= N\left(N-1\right)\int\prod_{i=3}^{N_{\uparrow}}\prod_{j=1}^{N_{\downarrow}}\mbox{d}\textbf{x}_i \mbox{d}\textbf{y}_j\left\vert\Psi_{GS}\left(\textbf{X}\right.\right.,&\left.\left.\textbf{Y},\textbf{x}_3,\cdots,\textbf{x}_{N_{\uparrow}},\textbf{y}_1,\cdots,\textbf{y}_{N_{\downarrow}};\,g_d\right)\right\vert^2+\\
&+\delta\left(\textbf{X}-\textbf{Y}\right)\rho_{GS}\left(\textbf{X};\,g_d\right)-\rho_{GS}\left(\textbf{X};\,g_d\right)\rho_{GS}\left(\textbf{Y};\,g_d\right),
\end{split}
\end{equation}
\begin{equation}\label{eq:funzionediscalamagnetization}
\begin{split}
\mathcal{M}(\textbf{x},\textbf{y};\,U_{r})=N_{\downarrow}N_{\uparrow}\int\prod_{i=2}^{N_{\uparrow}}\prod_{j=2}^{N_{\downarrow}}\mbox{d}\textbf{x}_i \mbox{d}\textbf{y}_j\left\vert\Psi_{GS}\left(\textbf{X},\textbf{x}_2,\cdots,\textbf{x}_{N_{\uparrow}},\textbf{Y},\textbf{x}_2,\cdots,\textbf{y}_{N_{\downarrow}};\,g_d\right)\right\vert^2 +\frac{N_{\downarrow}N_{\uparrow}}{N^2}\rho_{GS}\left(\textbf{X};\,g_d\right)\rho_{GS}\left(\textbf{Y};\,g_d\right),
\end{split}
\end{equation}
\end{widetext}
where $\Psi_{GS}\left(\textbf{x}_1,\cdots,\textbf{x}_{N_{\uparrow}},\textbf{y}_1,\cdots,\textbf{y}_{N_{\downarrow}};\,g_d\right)$ is the many-body ground-state wave-function and $\rho_{GS}\left(\textbf{x};\,g_d\right)$ is the one-particle density function of the $N$ body problem
\begin{equation}
\rho_{GS}\left(\textbf{x};\,g_d\right)=N\int\prod_{i=2}^{N}\,\left\vert\Psi_{GS}\left(\textbf{x},\textbf{x}_2\cdots,\textbf{x}_{N};\,g_d\right)\right\vert^{2},
\end{equation}
with the parameters $U_r$ and $g_d$ related by Eq. (\ref{eq: g-U}).\\

The rest of this paper is devoted to the analysis of the equations Eq.(\ref{eq:scalingrho}) - Eq.(\ref{eq:scalingmagnetization}).\\
\section{The two-body problem $\&$ the correspondence hypothesis}\label{sec:twobodyproblem}
In this section we discuss the \emph{correspondence hypothesis} in $d=1$ and $d=2$ for the two-body unpolarized problem ($N=2$ and $N_{\downarrow}=N_{\uparrow}=1$). The one- and two-dimensional cases are the only non-trivial cases. Indeed, both the continuum limit procedure and the RG analysis reported above suggest that the interparticle interaction becomes negligeble for $d \geq 3$, which means that the DR properties of the lattice problem in $d \geq 3$ are fully described by an hamiltonian of non-interacting fermions inside a harmonic trap.\\
\subsection{The ground-state wave-function of the unpolarized problem in $d=1$}\label{sec:oned}
According to the analysis in Sec.\ref{sec:regularization} the DR properties of the one-dimensional confined Hubbard model at very low temperature are in one-to-one correspondence respectively with the ground state-properties of the following many-body hamiltonians 
\begin{equation}\label{eq:continuum model dim 1}
\mathcal{H}_{c}^{(1)}=\sum_{i=1}^{N}\left(\frac{p_{i}^2}{2m}+\frac{1}{2}m\omega^2 x_{i}^{2}\right) +g_1\sum_{i=1}^{N_{\uparrow}}\sum_{j=1}^{N_{\downarrow}}\delta(x_i-x_j),
\end{equation}
which is the Gaudin-Yang model \cite{Gaudin,Yang} in the presence of an external harmonic potential.\\
This model and its properties have been considered and studied in several papars, with many different purposes (see for example Refs. \cite{Astrakharchik,Sowi,Volosniev,Grining,Bellotti}).
For a given number of particles $N$, the ground-state wave-function of the Hamiltonian model in Eq.(\ref{eq:continuum model dim 1}) can be found numerically for example by using methods Bethe-Ansatz formalism, DMRG or exact diagonalization. In our case, the explicit expression of all the eigenfunctions and their properties can be derived easily by using separation of variables and the results reported in Ref.\cite{delta1}. It turns out that for any value of the interaction strength $g_1$, the ground-state wave-function for the two-body unpolarized problem is always given by a $S=0$ state, where $S$ is the total spin of the two-body configuration. Therefore, the explicit expression of the ground-state wave function is 
\begin{equation}\label{eq:onedtwobodywavefunction}
\begin{split}
&\Psi_{GS}(x_1,\,x_2,\ S=0;\textbf{c})=\\
&=\beta\,\mbox{exp}\left(-\frac{x_1^2+x_2^2}{2\lambda^2}\right)U\left(-\frac{\nu(\alpha)}{2},\,\frac{1}{2};\,\frac{(x_1-x_2)^2}{2\lambda^2}\right)\chi_{S=0},\\
\end{split}
\end{equation}
where $\textbf{c}=(g_1,\,m,\,\omega)$, $\beta$ is a normalization constant, $\lambda=\sqrt{\frac{\hbar}{m\omega}}$ is the characteristic length scale of the harmonic oscillator, $\alpha=\frac{g_1}{\hbar\omega\lambda\sqrt{2}}$ and $U(a,\,b\,;x)$ is a \emph{Confluent Hypergeometric function} \cite{funzioni}. The parameter $\nu$, as discussed in details in \cite{delta1}, is related to the presence of the Dirac delta in the relative motion problem and it is related to the parameter $\alpha$ by the following equation
\begin{equation}
\nu\frac{\Gamma\left(\frac{1}{2}-\frac{\nu}{2}\right)}{\Gamma\left(1-\frac{\nu}{2}\right)}=\alpha,
\end{equation}
where $\Gamma(x)$ is the \emph{Euler Gamma function}.\\
\subsection{Trap-Size Scaling of the unpolarized problem in $d=1$}\label{sec:chmEgym}
\begin{figure}[t b]
\begin{center}
\subfigure[Interpolation for $U_r=10$]{\includegraphics[scale=0.6]{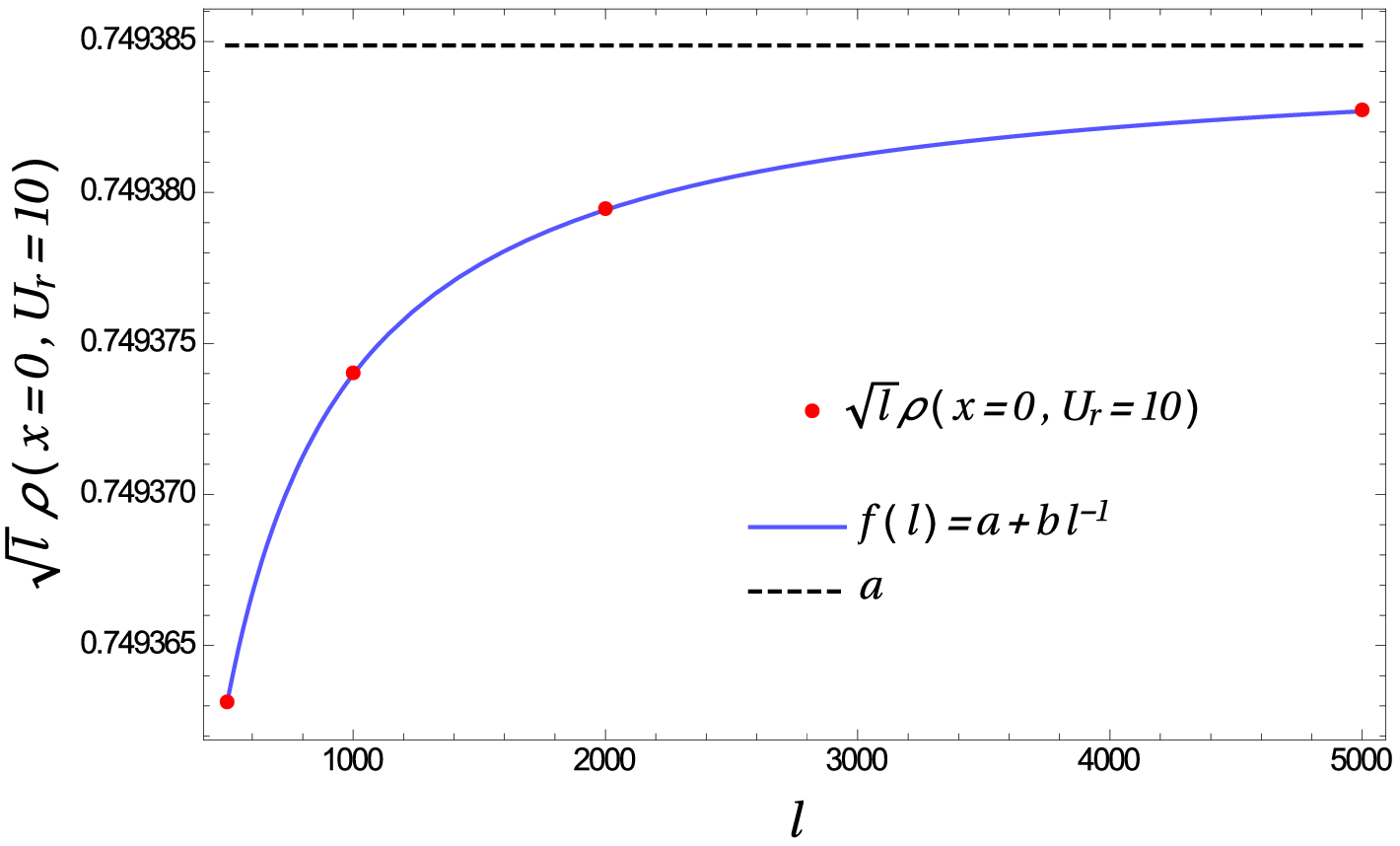}\label{fig:u10zero}}
\subfigure[Interpolation for $U_r=100$]{\includegraphics[scale=0.59]{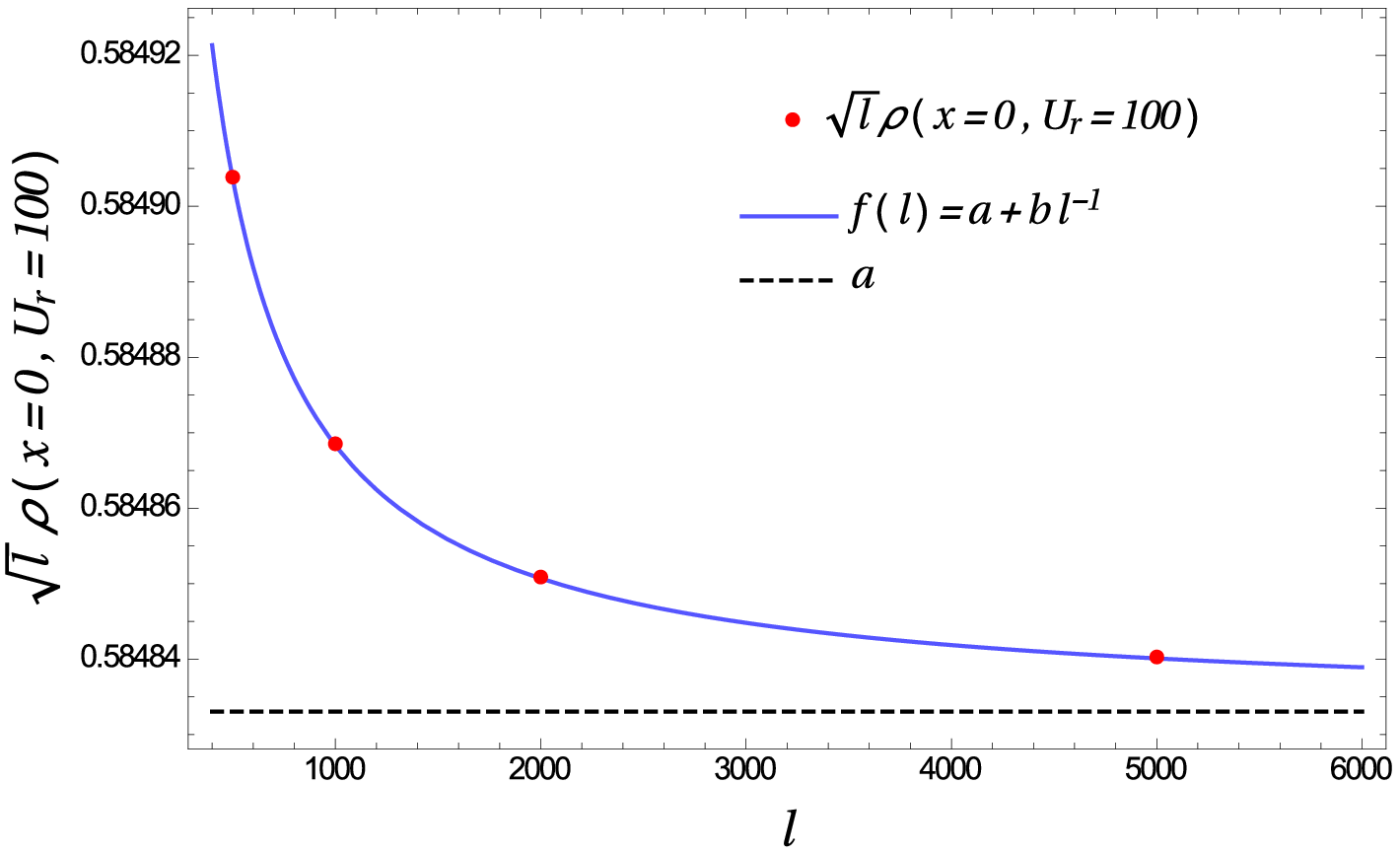}\label{fig:u100zero}}
\end{center}
\caption{Repulsive case: interpolation of $\sqrt{l}\rho(x=0,\,U_r=100)$ (\textbf{\ref{fig:u100zero}}) and $\sqrt{l}\rho(x=0,\,U_r=10)$ (\textbf{\ref{fig:u10zero}}) using the function $f(l)=a+b l^{-1}$ (solid blue line) to estimate the value of the parameter $a=\mathcal{R}(X=0,U_r)$ (dashed black line). The values of $\mathcal{R}(X=0,U_r=100)$ and $\mathcal{R}(X=0,U_r=10)$ have been reported in Table \ref{tab:confronto fra le densita nell'origine}.}\label{fig:repulsive case zero}
\end{figure}

\begin{figure}[t b]
\begin{center}
\subfigure[Interpolation for $U_r=-10$]{\includegraphics[scale=0.6]{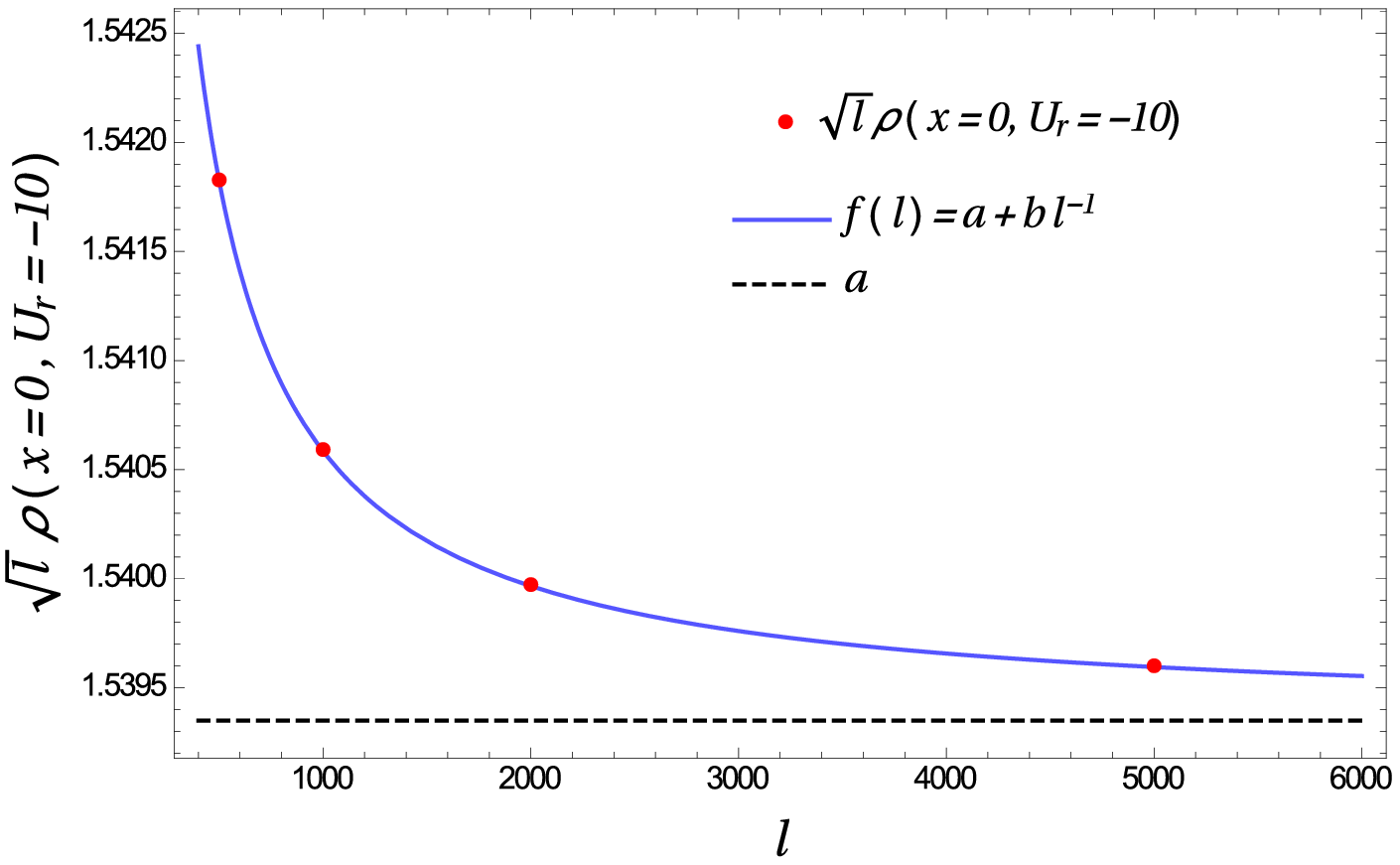}\label{fig:un10zero}}
\subfigure[Interpolation for $U_r=-100$]{\includegraphics[scale=0.575]{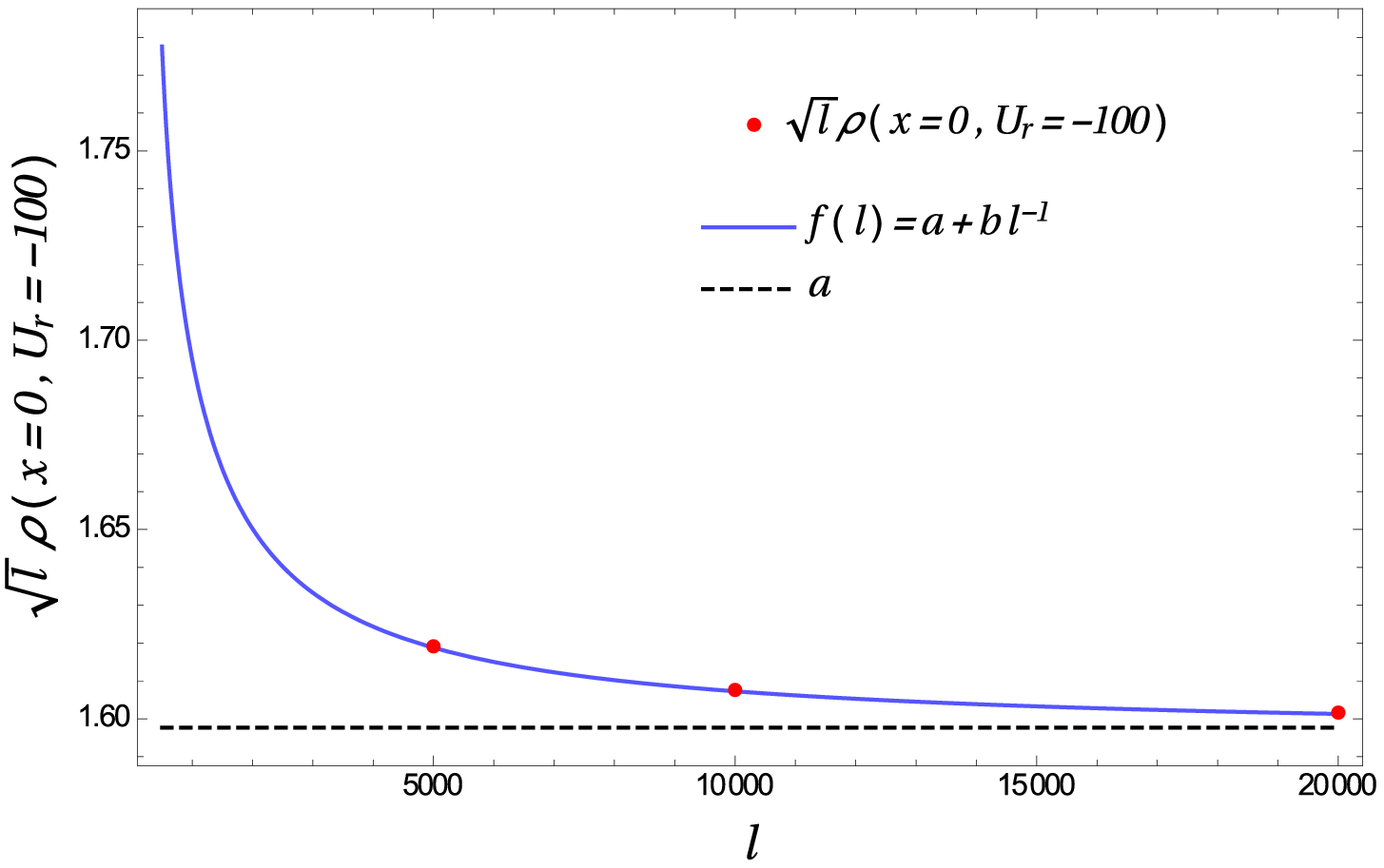}\label{fig:un100zero}}
\end{center}
\caption{Attractive case: Interpolation of $\sqrt{l}\rho(x=0,\,U_r=-10)$(\textbf{\ref{fig:un10zero}}) and $\sqrt{l}\rho(x=0,\,U_r=-100)$ (\textbf{\ref{fig:un100zero}}) using the function $f(l)=a+b l^{-1}$ (solid blue line) to estimate the value of the parameter $a=\mathcal{R}(X=0,U_r=10)$ (dashed black line). The values of $\mathcal{R}(X=0,U_r=-10)$ and $\mathcal{R}(X=0,U_r=-100)$ have been reported in Table \ref{tab:confronto fra le densita nell'origine}.}\label{fig:attractive case zero}
\end{figure}

\begin{figure}[h t b]
\includegraphics[width=.4\textwidth]{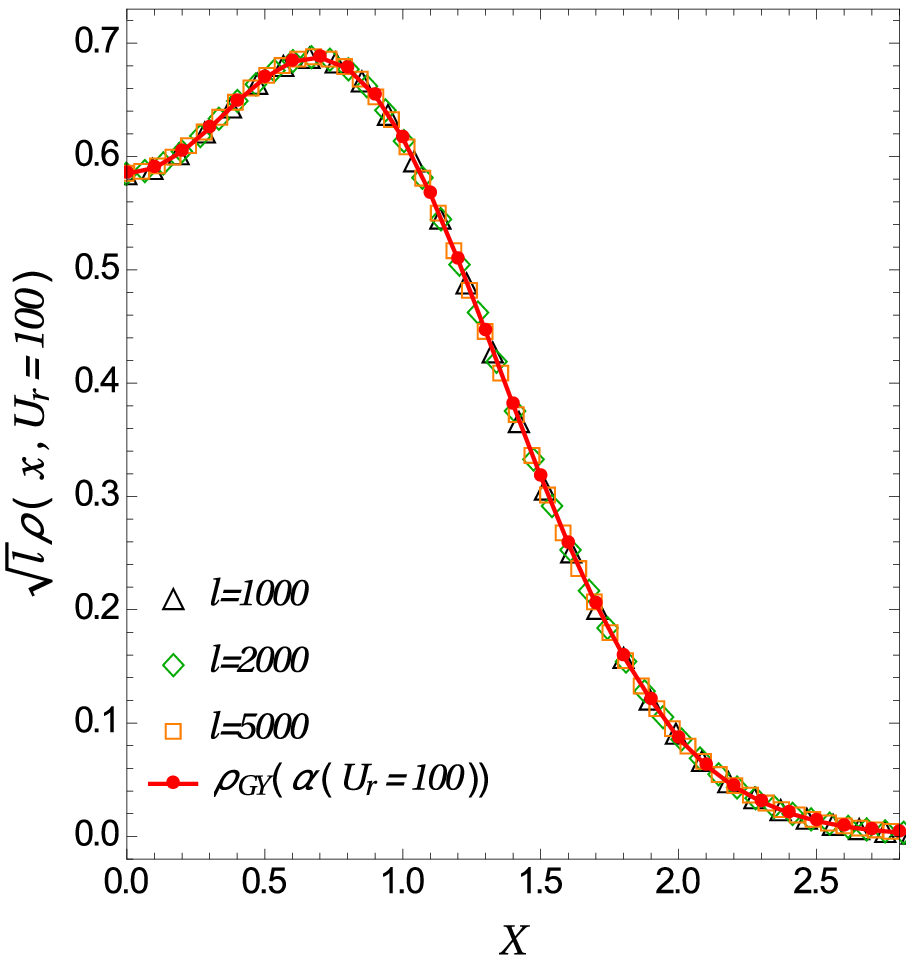}
\caption{TSS behaviour of $\sqrt{l}\rho(x,\,U_{r}=100)$  for different values of the trap-size $l$ and behaviour of the one-body density function $\rho_{GY}\left(X, \alpha(U_r)\right)$ (red dots).
}\label{fig:100}
\end{figure}

\begin{figure}[h t b]
\includegraphics[width=.4\textwidth]{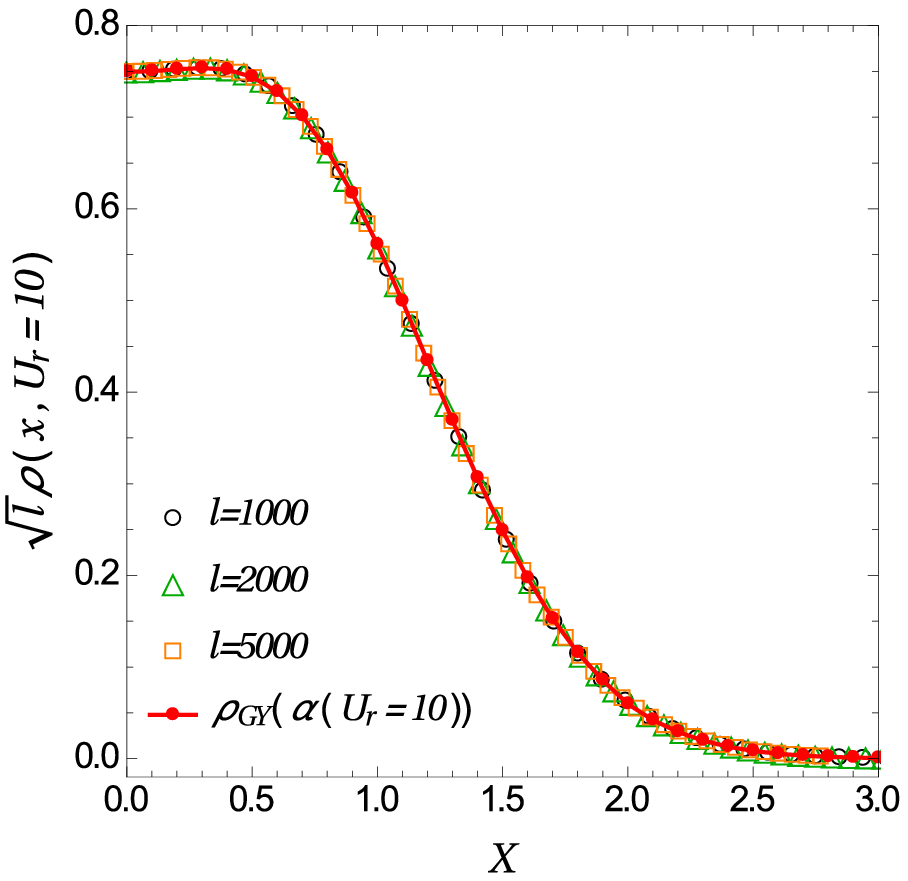}
\caption{TSS behaviour of $\sqrt{l}\rho(x,\,U_{r}=10)$ for different values of the trap-size $l$ and behaviour of the one-body density function $\rho_{GY}\left(X, \alpha(U_r)\right)$ (red dots).
}\label{fig:10}
\end{figure}

\begin{figure}[h t b]
\includegraphics[width=.4\textwidth]{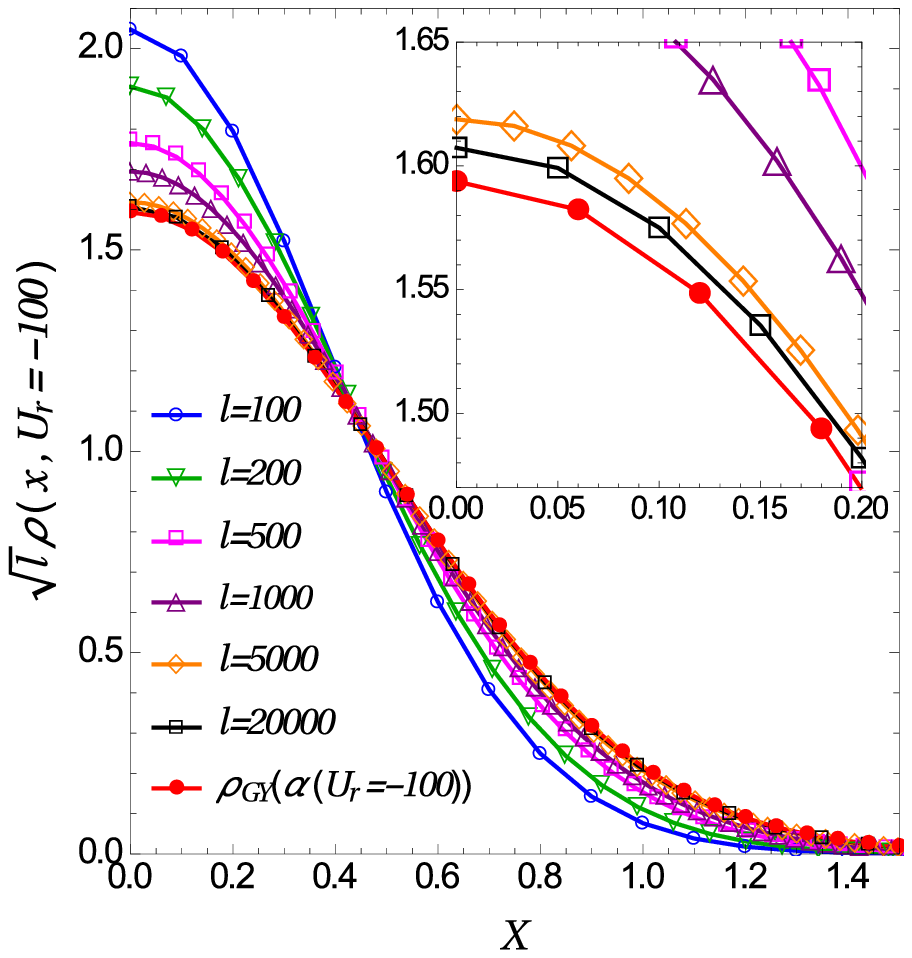}
\caption{TSS behaviour of $\sqrt{l}\rho(x,\,U_{r}=-100)$ for different values of the trap-size $l$ and behaviour of the one-body density function $\rho_{GY}\left(X, \alpha(U_r)\right)$ (red dots). 
}\label{fig:N100}
\end{figure}

\begin{figure}[h t b]
\includegraphics[width=.4\textwidth]{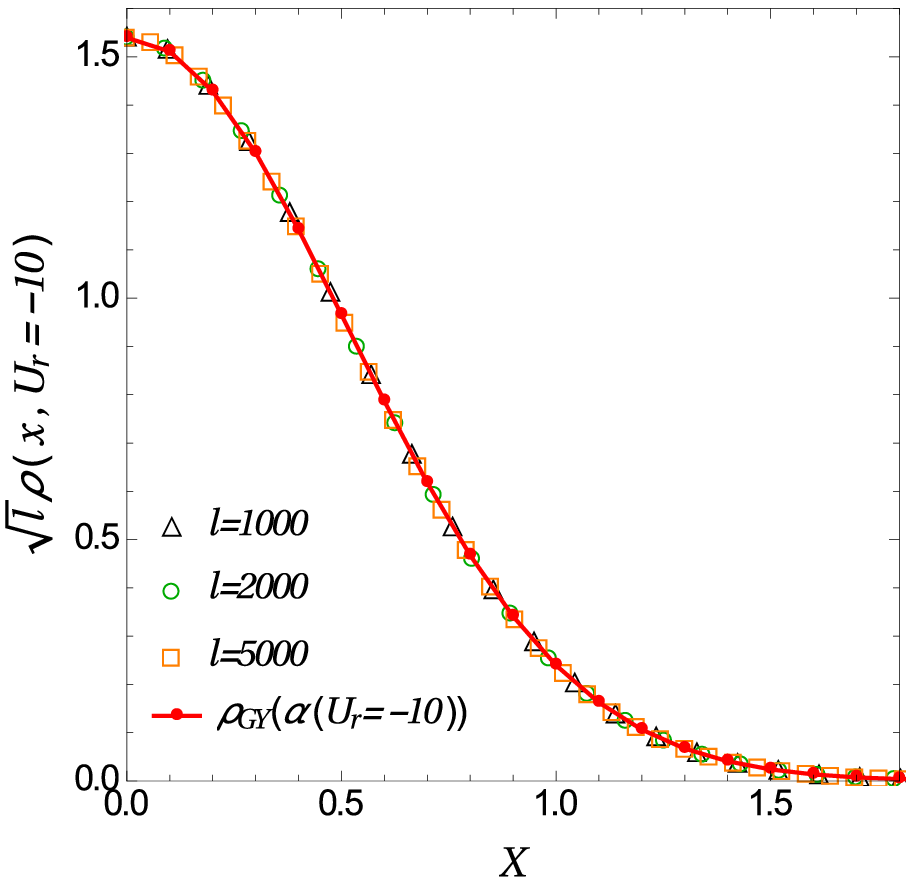}
\caption{TSS behaviour of $\sqrt{l}\rho(x,\,U_{r}=-10)$ for different values of the trap-size $l$ and behaviour of the one-body density function $\rho_{GY}\left(X, \alpha(U_r)\right)$ (red dots).
}\label{fig:N10}
\end{figure}

In $d=1$ the TSS behavior of the mean density at fixed $U_r$ is the following
\begin{equation}\label{eq:tssdim1}
\sqrt{l}\rho \left(x,\,U\right)=\mathcal{R}(X=x/\sqrt{l},U_r )+O(l^{-1}),
\end{equation}
where $O(l^{-1})$ indicates the finite $l$  corrections which disappear in the large $l$ limit \cite{TSS4}\\
The \emph{Correspondence Hypothesis} prescibes the scaling function in Eq.(\ref{eq:tssdim1}) to be \emph{exactly} the one-body density function associated to the ground-state wave-function of the continuum limit of the lattice problem, that is 
\begin{equation}\label{eq:andam asintotico uno d}
\mathcal{R}(X,\,U_r)=\rho_{GY}(X,\,g_1)
\end{equation}
where 
\begin{equation}
\rho_{GY}(X,\,g_1)=2\int\mbox{d}x\vert \Psi_{GS}(X,\,x,\ S=0;g_1,\,m,\,\omega)\vert^2,
\end{equation}
where the integration domain is the real axis.\\
Using adimensional variables, that is
\begin{equation}
\hbar=1,\,t=1,\,\lambda =1
\end{equation} 
the relation  between the $g_1$ and $U_r$ in Eq.(\ref{eq: g-U dim1}) becomes a relation between the parameter $\alpha$ and $U_r$. For the two-body unpolarized problem we have that these two parameters are related by the following law\\
\begin{equation}\label{eq:relazione di corrispondenza adimensionalizzata}
\alpha (U_r)=\frac{U_r}{2\sqrt{2}}
\end{equation}
We study the correspondence for different values of the rescaled coupling $U_r$ ($U_{r}=-100,-10,10,100$).\\
As a first proof of the correspondence we extrapolate $\mathcal{R}(X=0,\,U_r)$ and we compare this value with $\rho_{GY}(X=0,\,\alpha(U_r))$. The extrapolation of $\mathcal{R}(X=0,\,U_r)$ has been done by interpolation of the numerical results (exact diagonalization) obtained for finite values of the trap-size $l$. According to the scaling relation in Eq.(\ref{eq:tssdim1}), the large $l$ behaviour of $\sqrt{l}\rho(x=0,U_r)$ is the following
\begin{equation}
\sqrt{l}\rho(x=0,U_r)= a + b l^{-1}
\end{equation}
where the parameter $a=\mathcal{R}(X=0,\,U_r)$ and $b$ quantifies the entity of finite $l$ corrections. In Fig. \ref{fig:repulsive case zero} and Fig.\ref{fig:attractive case zero} have been reported the interpolating curves respectively for the \emph{repulsive} and the \emph{attractive} regimes. The results of the extrapolation procedure have been reported in Table \ref{tab:confronto fra le densita nell'origine}.\\
In Fig.\ref{fig:100} and in Fig.\ref{fig:10} we report the numerical results for the scaling function $\sqrt{l}\rho(x,\,U_{r})$ for $U_r=100$ and $U_r=10$. 
In Fig.\ref{fig:N100} and in Fig.\ref{fig:N10} we report instead the numerical results for the scaling function $\sqrt{l}\rho(x,\,U_{r})$  for $U_r=-100$ and $U_{r}=-10$ respectively. In Figs. \ref{fig:100}, \ref{fig:10}, \ref{fig:N100} and \ref{fig:N10} we also report the corresponding one-body density function $\rho_{GY}\left(x,\alpha(U_r)\right)$, where the values of $\alpha(U_r)$ have been chosen using the relation in Eq. (\ref{eq:relazione di corrispondenza adimensionalizzata}).\\
It is possible to observe that at increasing $l$, for all the values of the rescaled couplin $U_r$ considered, the data converge to the one-body density function $\rho_{GY}\left(x,\alpha(U_r)\right)$.
In addition, it interesting to note that the TSS behaviour reported here for the repulsive case is coherent with the experimental results reported in Ref. \cite{PhysRevLett.108.075303} where a system of two spin-$1/2$ $^6 Li$ atoms in a harmonic trap has been considered. At increasing interaction strength the energy of the two-body ground-stare increases and it reaches its maximum value when the coupling goes to $+\infty$. In this infinitely repulsive regime the $S=0$ and the $S=1$ configurations become degenerate in energy and the system of two spin-$1/2$ particles becomes a system of two indentical \emph{spinless} fermions. This phenomenon, which is the analogue of the \emph{fermionization} in one-dimensional systems of bosons \cite{Girardeau}, affects the one-body density. While increasing the interaction stregth from zero, two peaks placed symmetrically around $x=0$ appear in the one-body density and they become sharper as the coupling increases (compare Fig. \ref{fig:10} and Fig. \ref{fig:100}).\\

%

%
%
%
%
%

\begin{figure}[h]
\begin{center}
\includegraphics[scale=0.8]{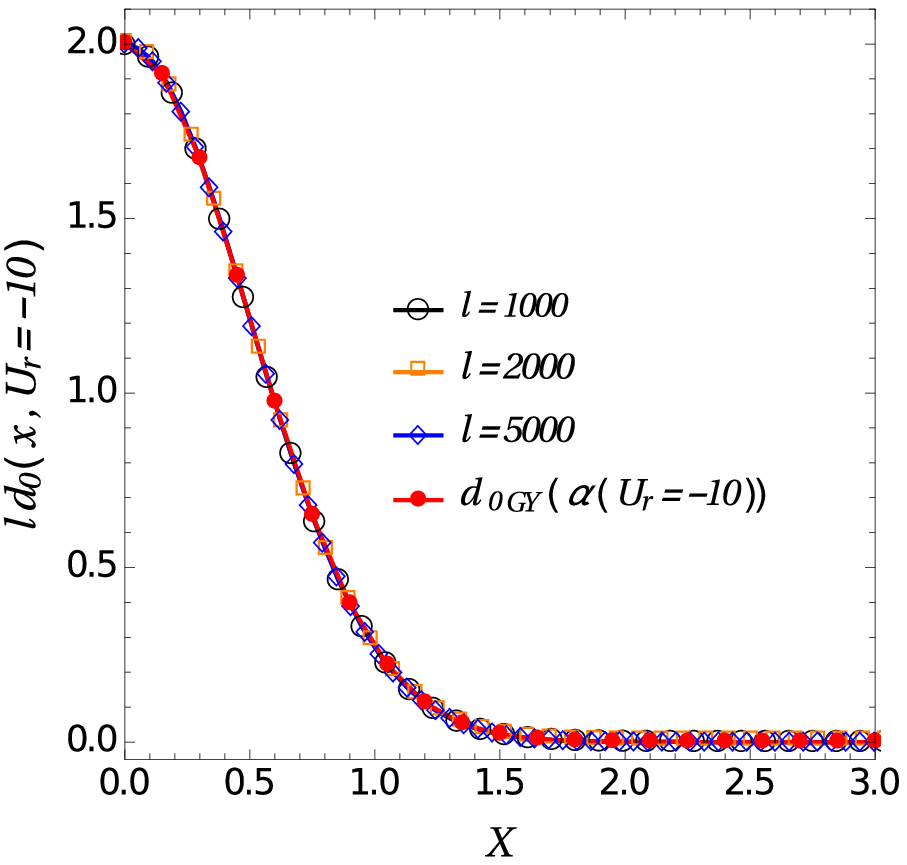}
\end{center}
\caption{TSS behaviour of the double-occupancy $l\,d_0(x,\,U_r)$ for different values of the trap-size $l$ ($l=1000,2000,5000$), compared with the scaling function $d_{0\,GY}(X\,;\alpha(U_r))$ of the two-body unpolarized problem for $U_{r}=-10$ (red dots).}\label{fig:tssDoubleOccupancyN10}
\end{figure}

\begin{figure}[h]
\begin{center}
\includegraphics[scale=0.8]{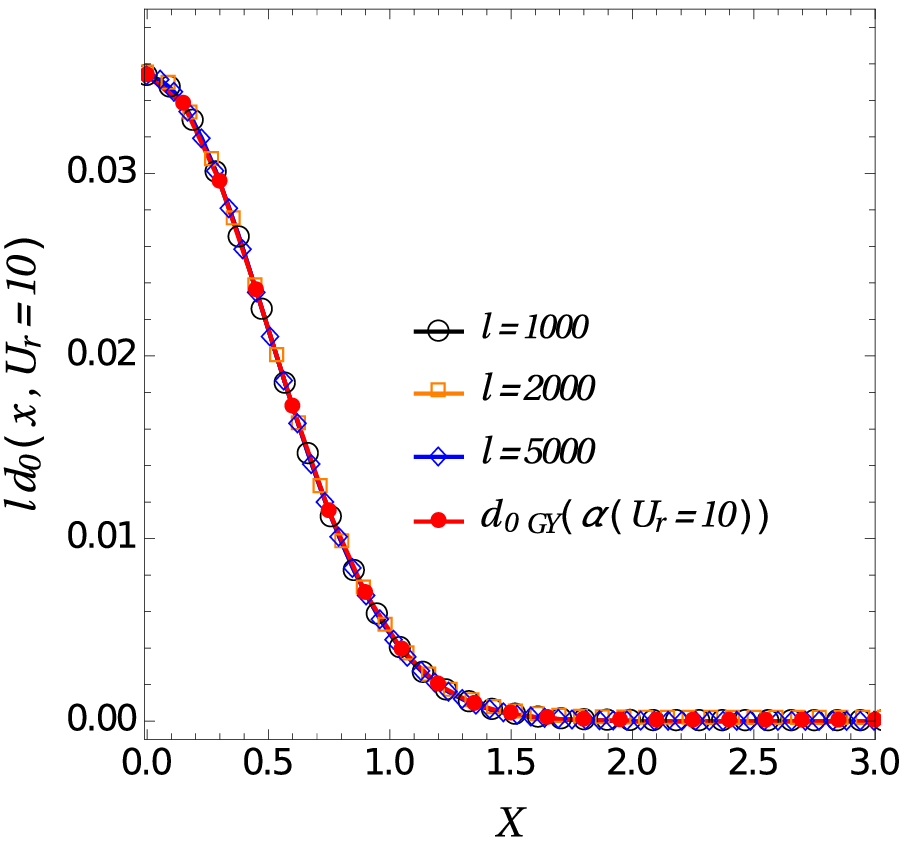}
\end{center}
\caption{TSS behaviour of the double-occupancy $l\,d_0(x,\,U_r)$ for different values of the trap-size $l$ ($l=1000,2000,5000$), compared with the scaling function $d_{0\,GY}(X\,;\alpha(U_r))$ of the two-body unpolarized problem for $U_{r}=10$ (red dots).}\label{fig:tssDoubleOccupancy10}
\end{figure}

\begin{figure}[h]
\begin{center}
\includegraphics[scale=0.8]{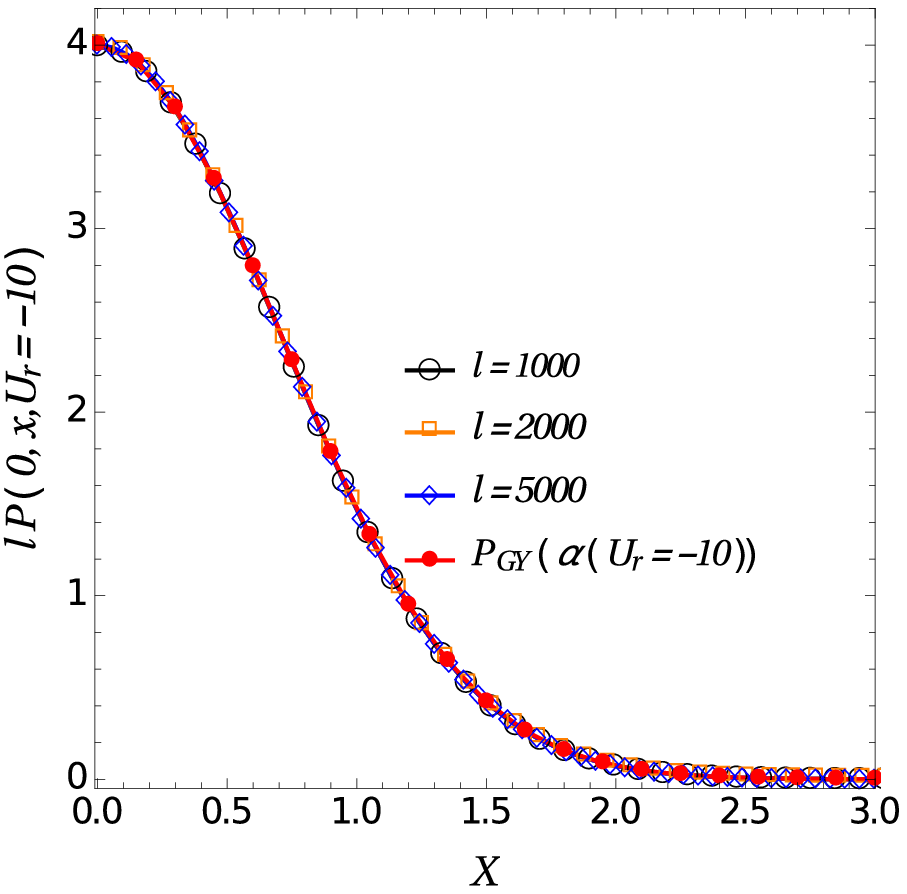}
\end{center}
\caption{TSS behaviour of the pair correlation $l\,P(0,\,x;\,U_r)$ for different values of the trap-size $l$ ($l=1000,2000,5000$) with a point fixed at the trap center, compared with the scaling function $P_{GY}(0,\,X\,;\alpha(U_r))$ of the two-body unpolarized problem for $U_{r}=-10$ (red dots). 
}\label{fig:tssPairCorrelationN10}
\end{figure}

\begin{figure}[h]
\begin{center}
\includegraphics[scale=0.8]{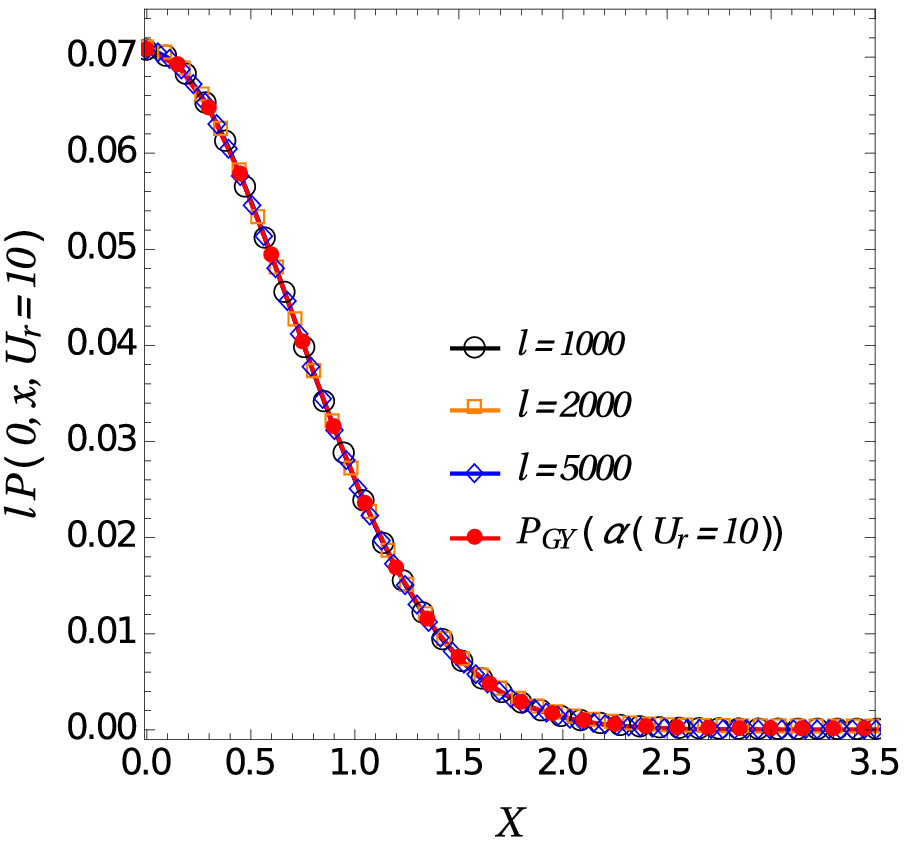}
\end{center}
\caption{TSS behaviour of the pair correlation $l\,P(0,\,x;\,U_r)$ for different values of the trap-size $l$ ($l=1000,2000,5000$) with a point fixed at the trap center, compared with the scaling function $P_{GY}(0,\,X\,;\alpha(U_r))$ of the two-body unpolarized problem for $U_{r}=10$ (red dots). 
}\label{fig:tssPairCorrelation10}
\end{figure}

\begin{figure}[h]
\begin{center}
\includegraphics[scale=0.8]{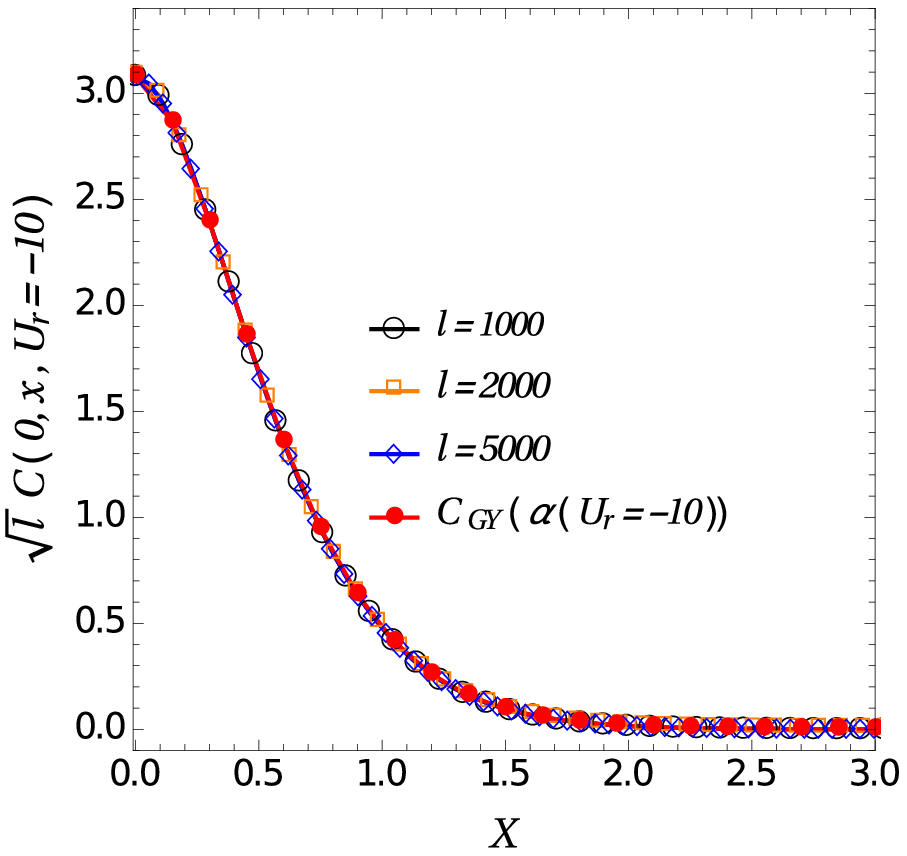}
\end{center}
\caption{TSS behaviour of the one-particle correlation  $\sqrt{l}\,C(0,\,x;\,U_r)$ for different values of the trap-size $l$ ($l=1000,2000,5000$) with a point fixed at the trap center, compared with the scaling function $C_{GY}(0,\,X\,;\alpha(U_r))$ of the two-body unpolarized problem for $U_{r}=-10$ (red dots).
}\label{fig:tssOneParticleCorrelationN10}
\end{figure}

\begin{figure}[h]
\begin{center}
\includegraphics[scale=0.8]{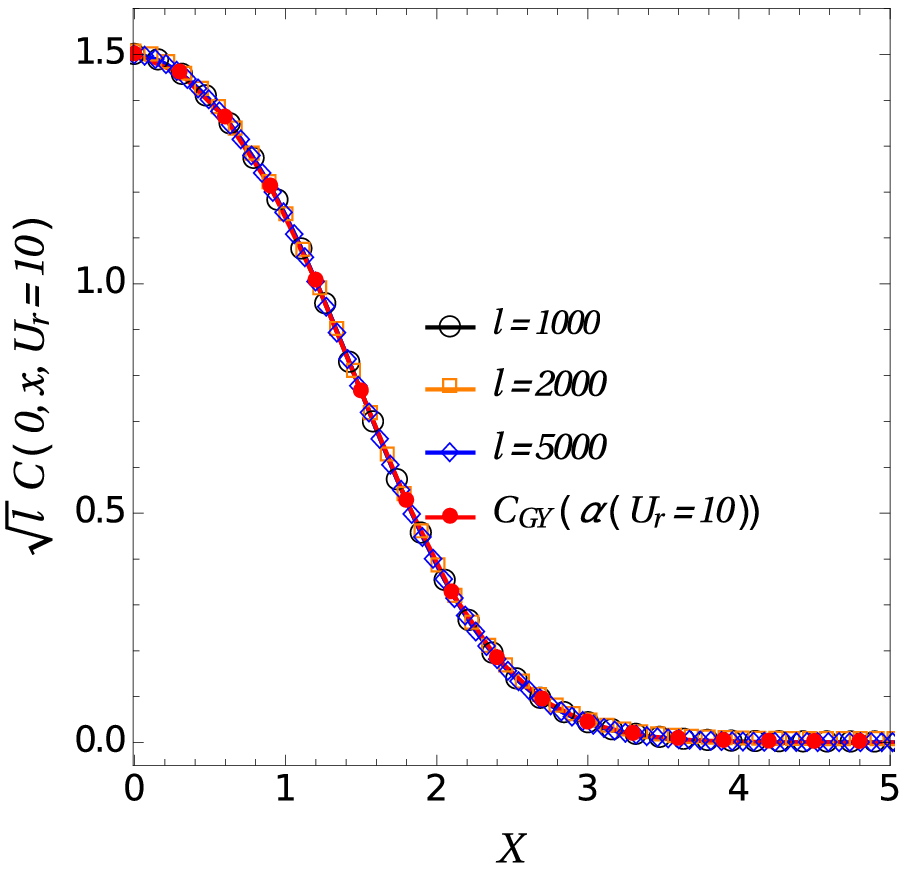}
\end{center}
\caption{TSS behaviour of the one-particle correlation  $\sqrt{l}\,C(0,\,x;\,U_r)$ for different values of the trap-size $l$ ($l=1000,2000,5000$) with a point fixed at the trap center, compared with the scaling function $C_{GY}(0,\,X\,;\alpha(U_r))$ of the two-body unpolarized problem for $U_{r}=10$ (red dots).
}\label{fig:tssOneParticleCorrelation10}
\end{figure}

\begin{figure}
\begin{center}
\includegraphics[scale=0.8]{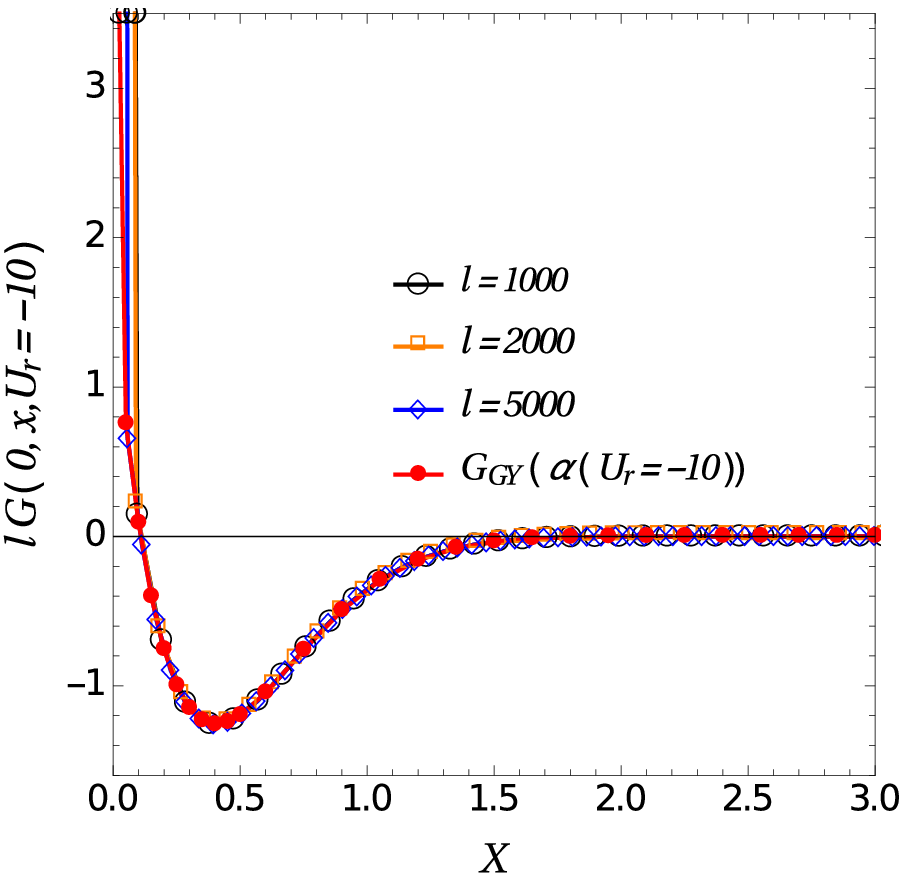}
\end{center}
\caption{TSS behaviour of the density-density correlation  $l\,G(0,\,x;\,U_r)$ for different values of the trap-size $l$ ($l=1000,2000,5000$) with a point fixed at the trap center, compared with the scaling function $G_{GY}(0,\,X\,;\alpha(U_r))$ of the two-body unpolarized problem for $U_{r}=-10$ (red dots).
}\label{fig:tssDensityDensityCorrelationN10}
\end{figure}

\begin{figure}
\begin{center}
\includegraphics[scale=0.8]{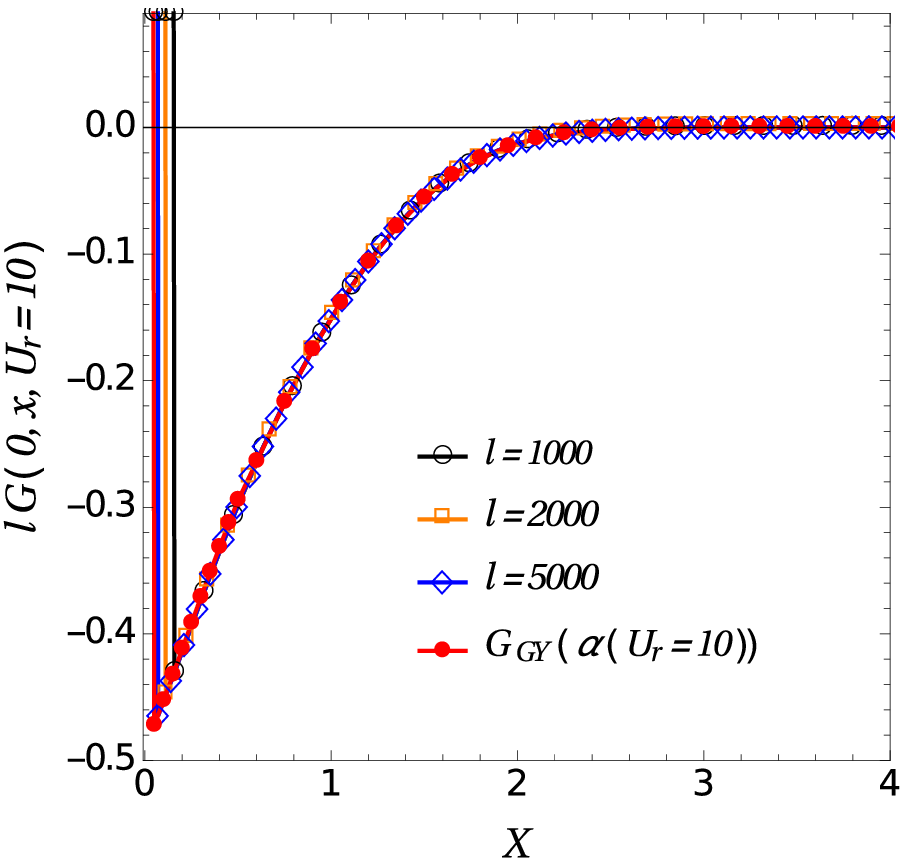}
\end{center}
\caption{TSS behaviour of the density-density correlation  $l\,G(0,\,x;\,U_r)$ for different values of the trap-size $l$ ($l=1000,2000,5000$) with a point fixed at the trap center, compared with the scaling function $G_{GY}(0,\,X\,;\alpha(U_r))$ of the two-body unpolarized problem for $U_{r}=10$ (red dots).
}\label{fig:tssDensityDensityCorrelation10}
\end{figure}

\begin{figure}
\begin{center}
\includegraphics[scale=0.8]{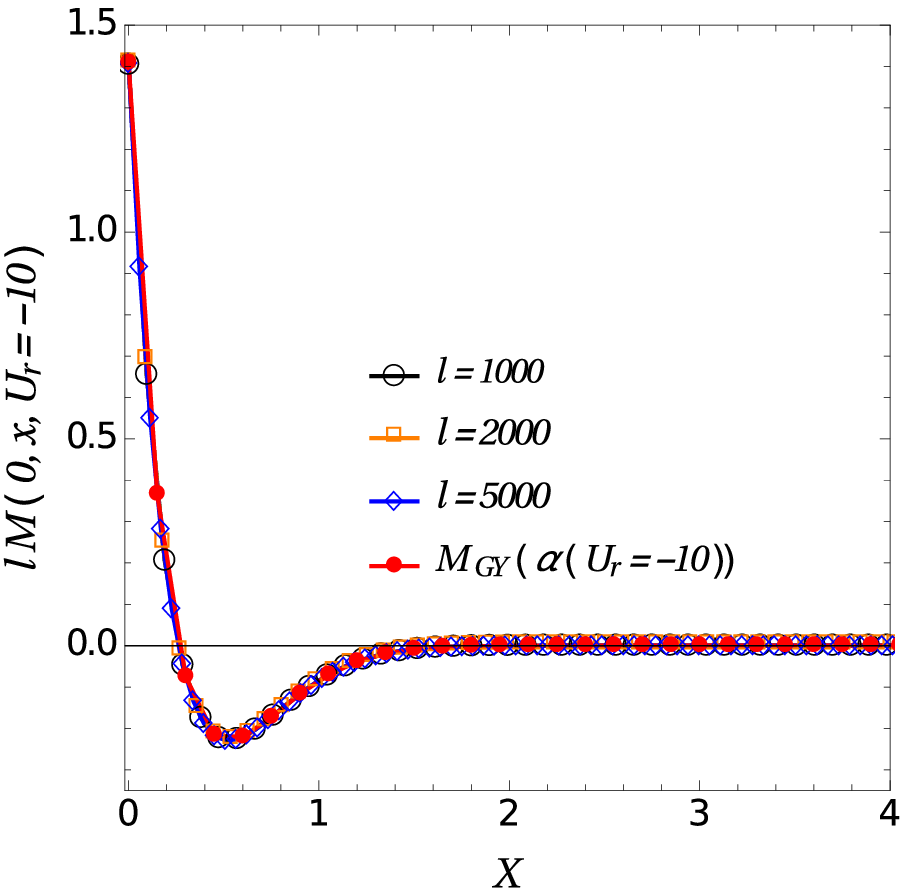}
\end{center}
\caption{TSS behaviour of the spin up - spin down density correlation $l\,M(0,\,x;\,U_r)$ for different values of the trap-size $l$ ($l=1000,2000,5000$) with a point fixed at the trap center, compared with the scaling function $M_{GY}(0,\,X\,;\alpha(U_r))$ of the two-body unpolarized problem for $U_{r}=-10$ (red dots).
}\label{fig:tssMagneticCorrelationN10}
\end{figure}

\begin{figure}
\begin{center}
\includegraphics[scale=0.8]{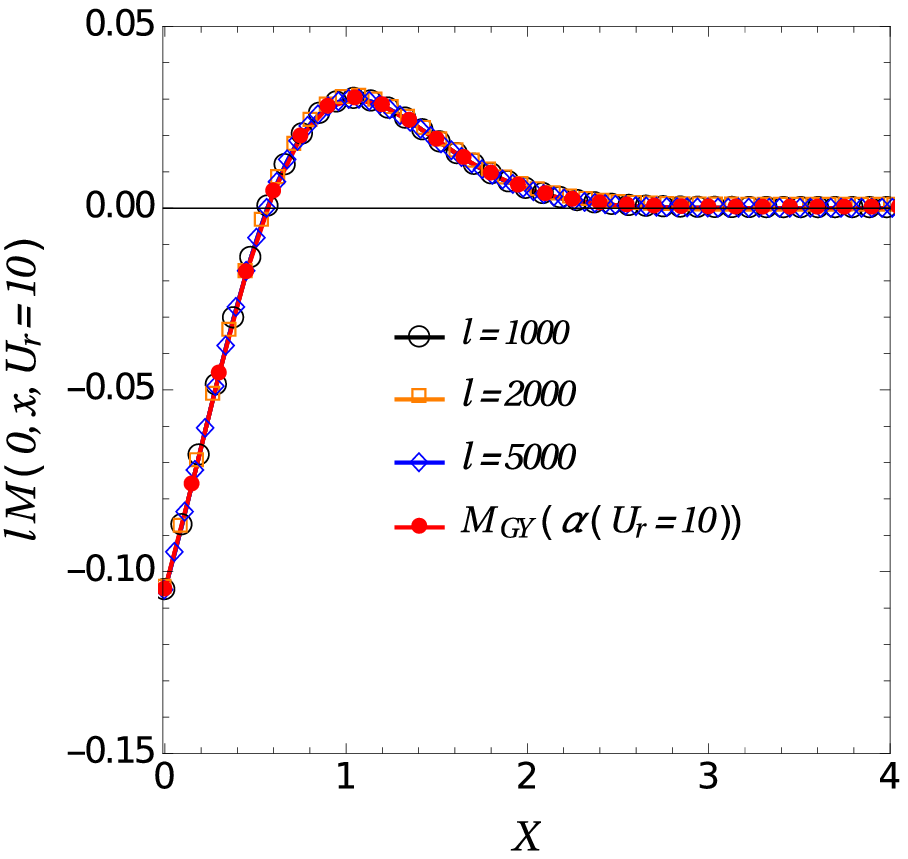}
\end{center}
\caption{TSS behaviour of the spin up-spin down density correlation $l\,M(0,\,x;\,U_r)$ for different values of the trap-size $l$ ($l=1000,2000,5000$) with a point fixed at the trap center, compared with the scaling function $M_{GY}(0,\,X\,;\alpha(U_r))$ of the two-body unpolarized problem for $U_{r}=10$ (red dots).
}\label{fig:tssMagneticCorrelation10}
\end{figure}

\begin{table}[h]
\begin{center}
\begin{tabular}{c c c }\hline
$U_r$ & $\mathcal{R}(X=0,\,U_r,)$ & $\rho_{GY}\left(X=0,\alpha\left(U_{r}\right)\right)$  \\
\hline\hline
100 & 0.58483303 & 0.58483303 \\
10& 0.74938486 & 0.74938486 \\
-10 & 1.53935 & 1.53935 \\
-100 & 1.59513 &  1.59513 \\
 \hline
\end{tabular}
\caption{Numerical estimates for the scaling function $\mathcal{R}(X=0,\,U_{r})$ and exact numerical values for the density function $\rho_{GY}\left(x=0,\alpha(U_r)\right)$ in correspondence of different values of the parameter $U_r$.
}\label{tab:confronto fra le densita nell'origine}
\end{center}
\end{table}
Let us now consider the TSS behaviour of the other observables listed in Sec. \ref{sec:correspondencehypothesis}. We study these observables for $U_{r}=-10$ and $U_{r}=10$. For the two-body unpolarized case the scaling functions reported in Eq. (\ref{eq:funzionediscaladouble}) - Eq. (\ref{eq:funzionediscalamagnetization}) reduce respectively to the following expressions:
\begin{equation}\label{eq:scalfundouble1d}
d_{0 GY}\left(x;\,g_d\right)=\left\vert\Psi_{GS}\left(x,x;\,g_1\right)\right\vert^{2},
\end{equation}
\begin{equation}\label{eq:scalfunpair1d}
P_{GY}\left(x;\,g_1\right)=\Psi_{GS}^{*}\left(x,x;\,g_1\right)\Psi_{GS}\left(y,y;\,g_1\right) + c\,.c\,. \,,
\end{equation}
\begin{equation}\label{eq:scalfunoneparticle1d}
\begin{split}
&C_{GY}\left(x,\,y;\,g_1\right)=\int \mbox{d}t\left[\Psi_{GS}^{*}\left(x,t;\,g_1\right)\Psi_{GS}\left(y,t;\,g_1\right)\right.+\\
&+\left.\Psi_{GS}^{*}\left(y,t;\,g_1\right)\Psi_{GS}\left(x,t;\,g_1\right) + c\,. \,c.\,\right],
\end{split}
\end{equation}
\begin{equation}\label{eq:scalfundensity1d}
\begin{split}
&G_{GY}\left(x,\,y;\,g_1\right)=\left\vert\Psi_{GS}\left(x,y;\,g_1\right)\right\vert^2+\left\vert\Psi_{GS}\left(y,x;\,g_1\right)\right\vert^2+\\
&+2\,\delta\left(x-y\right)\int\mbox{d}t\, \Psi_{GS}^{*}\left(x,t;\,g_1\right)\Psi_{GS}\left(y,t;\,g_1\right)-\\
&- \rho_{GY}\left(x;\,g_1\right)\,\rho_{GY}\left(y;\,g_1\right),
\end{split}
\end{equation}
\begin{equation}\label{eq:scalfunmagnetization1d}
\begin{split}
M_{GY}\left(x,\,y;\,g_1\right)=\left\vert\Psi_{GS}\right.&\left.\left(x,y;\,g_1\right)\right\vert^{2}-\\
&-\frac{1}{4}\rho_{GY}\left(x;\,g_1\right)\,\rho_{GY}\left(y;\,g_1\right).
\end{split}
\end{equation}
In Fig.(\ref{fig:tssDoubleOccupancyN10}) and Fig.(\ref{fig:tssDoubleOccupancy10}) we report the numerical results for the scaling function $l\, d_0(x;\,U_{r})$ for the attractive and repulsive regimes respectively. In Fig.(\ref{fig:tssPairCorrelationN10}) and Fig.(\ref{fig:tssPairCorrelation10}) we present results for the scaling behaviour of the pair-correlation function $l\, P(0,\,x;\,U_{r})$. The scaling behaviour of the one-particle correlation has been shown in Fig.(\ref{fig:tssOneParticleCorrelationN10}) and Fig.(\ref{fig:tssOneParticleCorrelation10}). The results for the density-density connected correlation and those for the correlation function between spin up and spin down densities are reported respectively in Fig.(\ref{fig:tssDensityDensityCorrelationN10})- Fig.(\ref{fig:tssDensityDensityCorrelation10}), and in Fig.(\ref{fig:tssMagneticCorrelationN10}) - Fig.(\ref{fig:tssMagneticCorrelation10}). In all these figures we report also the corresponding scaling function, whose explicit exressions are given in Eqs (\ref{eq:scalfundouble1d})-(\ref{eq:scalfunmagnetization1d}).\\
As it happens in the case of the mean-density, for all the observables considered in this work the data for different values of the trap-size clearly approach to the same scaling function. This function is the one prescribed by the continuum limit procedure discussed in Sec (\ref{subsec:continuumlimit}).
Therefore, the results shown in this section fully support the correspondence hypothesis in Eq.(\ref{eq:andam asintotico uno d}) and the correspondence between the two-body interaction strength prescribed by Eq.(\ref{eq:relazione di corrispondenza adimensionalizzata}). In other words our results prove that there is a one-to-one correspondence between the low density TSS properties of the confined Hubbard model and those of the continuous theory deduced in the Sec.\ref{subsec:continuumlimit} in $d=1$.\\
\subsection{The ground-state wave-function of the unpolarized problem in $d=2$}\label{sec:twod}
As discussed in Sec.\ref{subsec:continuumlimit}, the continuum limit procedure prescribes that in $d=2$ the confined-Hubbard model is in exact correspondence with the following many-body hamiltonian
\begin{equation}\label{eq:continuum model dim 2}
\mathcal{H}_{c}^{(2)}=\sum_{i=1}^{N}\left(\frac{\textbf{p}_{i}^2}{2m}+\frac{1}{2}m\omega^2 \textbf{x}_{i}^{2}\right) +g_2\sum_{i=1}^{N_{\uparrow}}\sum_{j=1}^{N_{\downarrow}}\delta(\textbf{x}_i-\textbf{x}_j).
\end{equation}
As for the $d=1$ case, we want to determine the ground-state wave-function of this hamiltonian. 
The main problem with this hamiltonian is that the Dirac delta in $d=2$ introduces divergences in wave functions and it is not possible to apply directly the method and the ideas used above for the $d=1$ case. One has first to remove these divergencies. This can be done by regularizing the two-body interaction. This regularization can be achieved by replacing the contact interaction in Eq.(\ref{eq:continuum model dim 2}) with the following contact interaction \cite{Busch1998,PhysRevA.74.022712}
\begin{equation}\label{eq:regularization of the interaction term}
V\left(\textbf{r}\right)=g_2\,\delta(\textbf{r})\left[1-\mbox{log}(r)r\frac{\partial}{\partial r}\right]
\end{equation}
where $\textbf{r}=\textbf{x}_1-\textbf{x}_2$ and $r=\vert\textbf{r}\vert$.\\
This version of the interaction term cancels the logarithmic divergence of the ground-state wave-function in correspondence of $r\to 0$ and it behaves exactly as the two-dimensional Dirac delta. As in $d=1$ the ground-state wave-function is a $S=0$ state and it has the following expression
\begin{equation}\label{eq:ground state for the 2d problem}
\begin{split}
&\Psi(\textbf{x}_1,\,\textbf{x}_2,\,S=0;\,\textbf{c})=\\
&=\beta\mbox{exp}\left(-\frac{\textbf{x}^2_1+\textbf{x}^2_2}{2\lambda^2}\right)U\left(\frac{\nu-1}{2},\,1,\,\frac{(\textbf{x}_1-\textbf{x}_2)^2}{2}\right),\\
\end{split}
\end{equation}
where $\textbf{c}=(g_2,\,m,\,\omega)$ is the set of parameters entering in Eq.(\ref{eq:continuum model dim 2}), $\beta$ is a normalization factor and $U(a,\,b;\,x)$ is again a confluent hypergeometric function.\\
As for the $d=1$ case, the parameter $\nu$ in Eq.(\ref{eq:ground state for the 2d problem}) is related to the presence of the contact interaction in the two-body problem and it is retaled to the interaction strength $g_2$ and all the other parametres entering in the hamiltonian model by the following relation \cite{Busch1998,PhysRevA.74.022712}
\begin{equation}\label{eq:generatingfunctionforthed2problem}
2\gamma -\mbox{log}(2)+\psi \left(\frac{\frac{1}{2}-\nu}{2}\right)=\frac{8\pi\hbar\omega}{g_{2}},
\end{equation}
where $\gamma$ is the \emph{Euler constant} and $\psi(x)$ the \emph{Digamma function} \cite{funzioni}.\\
In the following section we use the wave-function in Eq.(\ref{eq:ground state for the 2d problem}) to construct the one-body density function associated to the two-body problem. We compare this function to the numerical results obtained for the scaling function of the mean-density of the lattice problem to show that the correspondence holds also in $d=2$ (at least in the repulsive case).
\subsection{Trap-Size Scaling of the Ground-State Density and Analytical Solution for the unpolarized problem in $d=2$}
\begin{figure}[h]
\begin{center}
\includegraphics[scale=0.8]{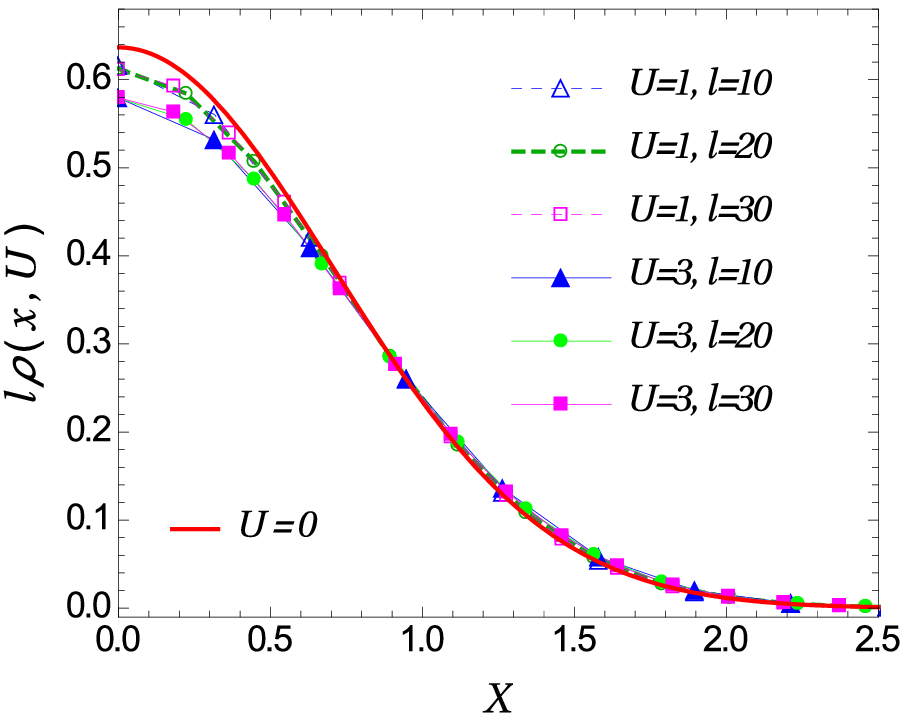}
\end{center}
\caption{
Numerical results obtained for the rescaled density function $l\rho(\textbf{x},\,U)$ compared with the one-body density function of the two-dimensional non-interacting problem ($U=0$, red solid line).
}\label{fig:tss2dplusNonint}
\end{figure}

\begin{figure}[h t b]
\begin{center}
\includegraphics[scale=0.8]{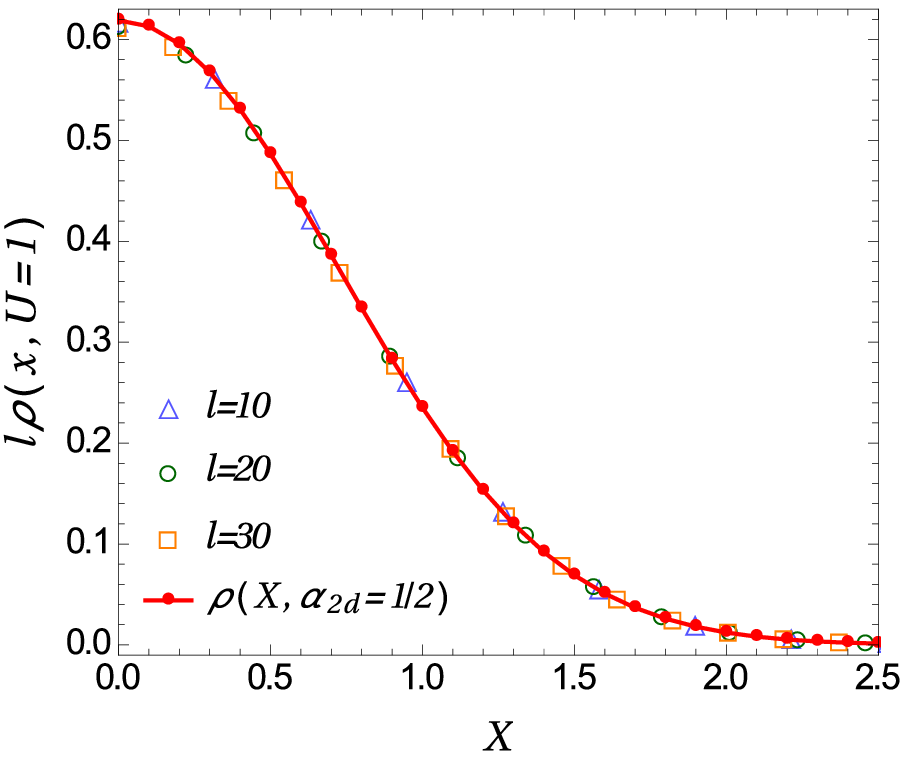}
\caption{Numerical results for the rescaled-density $l\rho(\textbf{x},\,U=1)$ ($l=10,20,30$) compared with the one-body density associated to the ground-state wave-function in correspondence of $\alpha_{2d}=1/2$.}\label{fig:TSS2danlitical1}
\end{center}
\end{figure}

\begin{figure}[h t b]
\begin{center}
\includegraphics[scale=0.8]{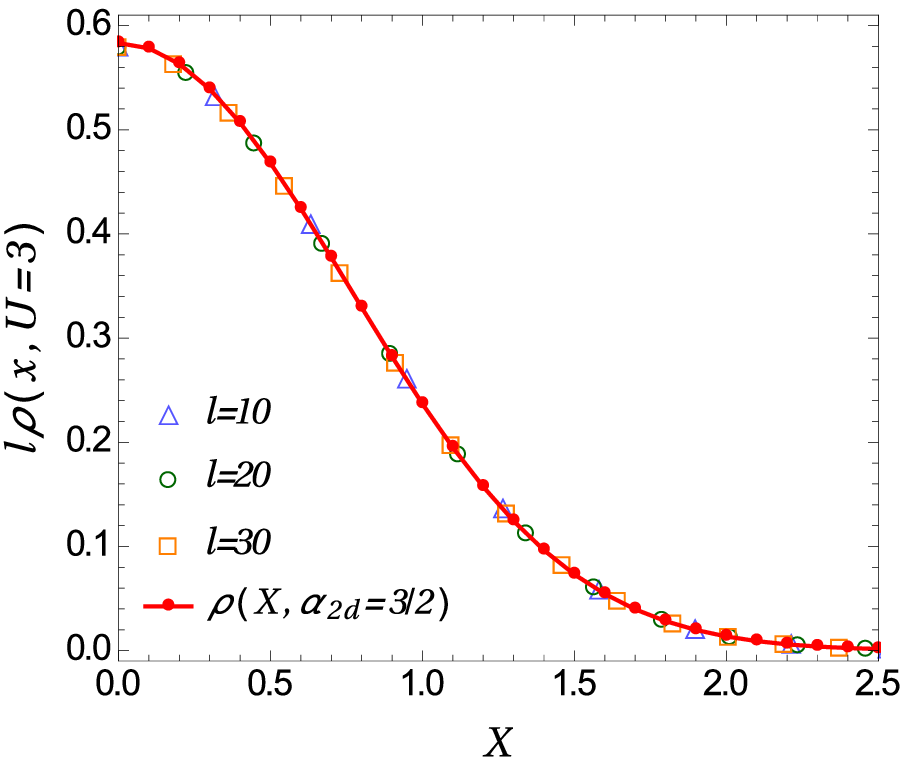}
\caption{Numerical results for the rescaled-density $l\rho(\textbf{x},\,U=3)$ ($l=10,20,30$) compared with the one-body density associated to the ground-state wave-function in correspondence of $\alpha_{2d}=3/2$.}\label{fig:TSS2danlitical2}
\end{center}
\end{figure}

In $d=2$, as discussed in Sec.\ref{sec:correspondencehypothesis}, the TSS behaviour of the particle density under a scaling transformation of the lattice position $\textbf{x}$ is the following 
\begin{equation}\label{eq:tss2d}
l\rho(\textbf{x},U)=\mathcal{R}^{(2d)}\left(\textbf{X},\,U\right)+O(l^{0}),
\end{equation}
where $\mathcal{R}^{(2d)}\left(\textbf{X},\,U\right)$ and $O(l^{0})$ denote respectively the scaling function of the $d=2$ problem and the finite-$l$ corrections.\\
As in the $d=1$ case we compare the behaviour of rescaled density in Eq.(\ref{eq:tss2d}) at increasing $l$ with the one-body density associated to the ground-state wave-function reported in Eq.(\ref{eq:ground state for the 2d problem}) assuming the correspondence relation reported in Eq.(\ref{eq: g-U dim2}) to be true.\\
As in the $d=1$ case we can introduce the one-body density function associated to the wave-function $\Psi$ in Eq.(\ref{eq:ground state for the 2d problem})
\begin{equation}\label{eq:ground-statedensity}
\begin{split}
&\rho(\textbf{x}_1,\, g_{2})=2\int \mbox{d}\textbf{x}_2\,\vert\Psi(\textbf{x}_1,\,\textbf{x}_2,\,S=0;\,\textbf{c})\vert^2=\\
&=2\int \mbox{d}\textbf{x}_2\,\beta^2\mbox{exp}\left[-\left(\frac{\textbf{x}_1^2+\textbf{x}_2^2}{\lambda^2}\right)\right] \times \\
&\quad\quad\quad\times U^2\left(\frac{\nu-1}{2},\,1,\,\frac{\left(\textbf{x}_1-\textbf{x}_2\right)^2}{2\lambda^2}\right),\\
\end{split}
\end{equation}
where the integration domain is $\mathbb{R}^2$.\\
By inspection of Eq.(\ref{eq:ground state for the 2d problem}), it is easy to see that the one-body density function is invariant under rotations of the vector $\textbf{x}_1$. This means that this one-body density is completely characterized by the knowledge of the $\rho(x,\, g_{2})\equiv\rho(\bar{\textbf{x}}_1,\, g_{2})$, with $\bar{\textbf{x}}_1=(x,\,0)$ and $x\geq 0$.\\
Using adimensional variables, that is 
\begin{equation}
\hbar=1,\quad t=1,\quad \lambda =1,
\end{equation}
the correspondence relation between the $g_{2}$ and the on-site coupling constant $U$ becomes 
\begin{equation}\label{eq:correspondencerelation2d}
\alpha_{2d}(U)=\frac{U}{2},
\end{equation}
where $\alpha_{2d}$ is the adimensional analogue of the interaction strength $g_{2}$.\\
As in the $d=1$ case, see Fig.\ref{fig:tss2dplusNonint}, while increasing $U$, the value of the rescaled density in $X=0$ decreases with respect to the non-interacting case and in addition numerical results appear to converge to a non-trivial curve at increasing $l$. This curve depends on the particular value of the on-site coupling $U$.\\
Assuming the correspondence relation in Eq.(\ref{eq:correspondencerelation2d}) to be true we compared the numerical results obtained for $l\rho(\textbf{x},\,U)$ for the two values $U=1$ and $U=3$ with the profiles of the one-body density functions associated to the wave-function in Eq.(\ref{eq:ground state for the 2d problem}) for the values of $\alpha_{2d}$ prescribed by Eq.(\ref{eq:correspondencerelation2d}). Results have been reported in Fig.\ref{fig:TSS2danlitical1} and Fig.\ref{fig:TSS2danlitical2}.\\
Results reported in Fig.\ref{fig:TSS2danlitical1} and in Fig.\ref{fig:TSS2danlitical2} support the correspondence hypothesis for $U>0$. Further studies are needed to prove explicitly the validity of the correspondence hypothesis for $U<0$.\\

\section{Summary and Conclusions}\label{sec:conclusions}
In the present work we have considered the $d$-dimensional Hubbard model in the presence of an external harmonic confinement coupled to the particle density. In particular we have considered this system in the so called \emph{Dilute Regime} (DR), which is the low-density regime achieved considering a fixed number of particles $N$ in the presence of a weak confinement. The ground-state properties of the one-dimensional case in the presence of an harmonic confinement have been extensively studied in \cite{TSS4} using both Renormalization Group (RG) techniques and the Trap-Size Scaling formalism (TSS). In particular in \cite{TSS4}, the authors argue that the leading TSS behaviour of the particle density reproduces the ground-state properties of a one-dimensional gas of spin-$1/2$ fermions described by the Gaudin-Yang model in the presence of an harmonic trap (GY model), with an interaction strength $g$ between particles with opposite polarization that is proportional to the variable $U_r=Ul^{1/2}$.\\
In the present work we have derived the continuous theory which describes the TSS properties of the Hubbard model in the presence of an external harmonic confinement in \emph{any} $d$. The expression of this continuous theory depends explicitly on the dimension $d$ of the lattice problem and its properties are in agreement with the RG analysis reported in \cite{TSS4}.\\
In $d=1$ the continuum limit of the lattice problem with $N$ particles is the GY model with the same number of particles. The correspondence hypothesis between the leading TSS behaviour of the particle density and the profiles associated to the one-body density function of the ground-state of the GY model has been tested comparing the numerical results obtained from the exact diagonalization of the lattice problem and the analytical solutions of the two-body unpolarized problem ($N=2$ and $N_{\uparrow}=N_{\downarrow}=1$). In particular we have found the analytic relation between on-site coupling $U$ of the Hubbard model and the interaction strength $g_1$ of the Gaudin-Yang model in presence of an external harmonic confinement. We have to stress that the relation which prescribes the correspondence between $g_1$ and $U$ does not depend on the number of particles $N$, so the its validity holds for any $N$ (see Eq.(\ref{eq: g-U dim1})).\\
In $d=2$ the continuum limit of the lattice problem is an interacting theory with a local interaction between particles with opposite spin which is shaped as a Dirac delta. The main problem with interactions of this kind in $d=2$ is that they have to be regularized to cancel the logarithmic divergences of the wave-functions. The regularization has been done following the same ideas reported in \cite{PhysRevA.74.022712}. As in $d=1$ we tested the correspondence hypothesis comparing numerical results and analytical results obtained for the ground-state configuration of the two-body unpolarized problem. Results reported support the correspondence between the DR properties of the two-dimensional confined Hubbard model and those of the continuum limit derived in Sec.\ref{sec:regularization}, in particular they confirm the validity of analytical relation reported in Eq. (\ref{eq: g-U dim2}) between the on-site interaction $U$ and the coupling constant $g_2$ (at least for $U\geq 0$).\\
In $d\geq 3$ results obtained performing the continuum limit confirm that the leading TSS behaviour of the Hubbard model in the DR is described by a non-interacting theory as obtained in \cite{TSS4} using RG methods.\\

\section*{Acknowledgements}
I wish to thank E. Vicari and D. Rossini for helpful discussions and M. Campostrini for the support with the numerical analysis.




\end{document}